\documentclass[preprint]{revtex4}

\usepackage{amsmath, amssymb}
\usepackage{natbib}
\usepackage[colorlinks, 
linkcolor=blue,
urlcolor=blue,
citecolor=blue,
bookmarksopen=false]{hyperref}

\usepackage{amsmath}
\usepackage{graphicx, xcolor}
\usepackage{caption}
\usepackage{amsmath}
\usepackage{float}

\def\iotabar{\lower3pt\hbox{$\mathchar'26$}\mkern-7mu\iota}
\newcommand {\aplt} {\ {\raise-.5ex\hbox{$\buildrel<\over\sim$}}\ }

\newcommand{\dotcross}{ \raise 0.65ex\hbox{${\scriptstyle {{_{\displaystyle \cdot}}\atop\times}}$} }
\newcommand{\crossdot}{ \raise 0.5ex\hbox{${\scriptstyle {{_\times}\atop{\displaystyle \cdot}}}$} }

\newcommand{\sumsig}{ \raise -1.3ex\hbox{${{\displaystyle \sum}\atop{\scriptstyle \sigma}}$} }
\newcounter{appnumb}

\begin{document}

\title{Exploration of the parameter space of piecewise omnigenous stellarator magnetic fields}
\author{J.~L. Velasco, E. S\'anchez, I. Calvo}
\affiliation{Laboratorio Nacional de Fusi\'on, CIEMAT, 28040, Madrid, Spain}
\email{joseluis.velasco@ciemat.es}\homepage{\\ http://fusionsites.ciemat.es/jlvelasco}

\date{\today}

\begin{abstract}
  
Piecewise omnigenous fields are stellarator magnetic fields that are optimized with respect to radial neoclassical transport thanks to a second adiabatic invariant that is piecewisely constant on the flux-surface. They are qualitatively different from omnigenous fields (including quasi-isodynamic or quasisymmetric fields), for which the second adiabatic invariant is a flux-surface constant. Piecewise omnigenous fields thus open an alternative path towards stellarator reactors. In this work, piecewise omnigenous fields are characterized and parametrized in a systematic manner. This is a step towards including piecewise omnigenity as an explicit design criterion in stellarator optimization, and towards a systematic study of the properties of nearly piecewise omnigenous stellarator configurations.
    
\end{abstract}

\maketitle

\section{Introduction}\label{SEC_INTRO}

Stellarator fields need to be optimized in order to be candidates for a fusion reactor based on magnetic confinement. In a generic stellarator, neoclassical transport, i.e., the transport caused by the combination of the inhomogeneity of the magnetic field and particle collisions, is unacceptably large in low-collisionality plasmas. This is in contrast with tokamak fields, whose axisymmetry yields small neoclassical transport at low collisionality. On the other hand, once optimized and built, stellarators have the advantage of being easier to operate, since their magnetic configuration is created by external coils, and they are therefore inherently free from current-driven instabilities.

The need for optimization stems from the existence of trapped particles, i.e., particles whose component of the velocity parallel to the magnetic field $\mathbf{B}$ vanishes along their trajectory (in the absence of collisions, passing particles are confined in any magnetic field with nested toroidal flux surfaces). Trapped particles  bounce back and forth along the magnetic field line while they slowly drift perpendicularly to it, either in the radial (perpendicularly to the flux surfaces) or the tangential direction (within the flux surface). In tokamaks, axisymmetry guarantees that the radial drift vanishes on average and trapped particles are confined in the absence of collisions. In stellarators, confining collisionless trapped-particle orbits requires a careful design of the magnetic field. Stellarators in which the radial magnetic drift averages to zero for all trapped particles are called omnigenous \cite{cary1997omni}. Omnigenity imposes severe constraints on the variation of the magnetic field strength on the flux-surface and, in particular, on the topology of the contours of constant magnetic field strength $B$, which have to close in the toroidal, helical or poloidal direction \cite{cary1997omni,parra2015omni}. Most new stellarator configurations are optimized to approach omnigenity \cite{henneberg2019qa,plunk2019direct,kinoshita2019cfqs,bader2020wistell,landreman2022preciseQS,landreman2022mapping,camachomata2022direct,jorge2022qi,sanchez2023qi,goodman2023qi,dudt2023omni}. However, this can lead to exceedingly complicated plasma shapes and coils \cite{strykowsky2009ncsx}. 

In~\cite{velasco2024pwO}, the concept of piecewise omnigenity has been introduced. Piecewise omnigenous (pwO) fields constitute a new family of stellarator fields with reduced transport in the deleterious $1/\nu$ neoclassical regime even though their $B$-contours do not have the topology required by omnigenity. This result has provided a framework that explains how fields with very small neoclassical transport (and other reactor-relevant properties) can exist very far from the usual definition of omnigenity. For instance, the standard configuration of Wendelstein 7-X (sometimes referred to as W7-X standard), the inward-shifted configuration of the Large Helical Device (LHD)~\cite{beidler2011ICNTS}, and configuration A of~\cite{bindel2023direct} have been argued in \cite{velasco2024pwO} to be relatively close to piecewise omnigenity. The first two configurations have been fundamental for obtaining record stellarator experimental plasmas \cite{yamada2005taue,beidler2021nature} and are consistent with reactor scenarios, see e.g.~\cite{miyazawa2014ffhrd1}.  The latter has even better neoclassical transport properties, according to numerical predictions.

Furthermore, the notion of piecewise omnigenous field introduced in~\cite{velasco2024pwO} may radically expand the range of configurations that can be candidates for fusion reactors, some of which might be easier to design, build or operate. Some of these configurations present interesting properties that could be general features of nearly pwO fields, such as a lower ratio of fast ion energy loss over fast ion particle loss, see the Supplemental Material of~\cite{velasco2024pwO}. This could enable efficient alpha heating with low helium ash accumulation. Other properties are likely specific of a subset of these configurations. For instance, ion temperature-gradient turbulence is mild in LHD as compared with W7-X~\cite{regana2021imp3d,thienpondt2024comp}. A thorough study would be required to identify configurations (or families of configurations) whose properties, beyond neoclassical transport, make them good candidates for a fusion reactor.

A possible starting point for this exercise would be a characterization of the existing types of pwO fields through the main features of the variation of $B$ on the flux-surface. In the case of omnigenous fields, \textit{helicity} is the most salient property of $B(\theta,\zeta)$, where $\theta$ and $\zeta$ are the toroidal and poloidal Boozer angles. In an omnigenous field, the maxima of $B$, $B_\mathrm{max}$, are straight lines of constant $M\theta-N_pN\zeta$, where $N$ and $M$ are integers and $N_p$ is the number of field periods; the rest of $B$-contours close in the same direction, toroidally ($N=0$), poloidally ($M=0$) or helically ($M\neq 0$, $N\neq 0$). A quasi-isodynamic (QI) field is an omnigenous field in which $M=0$. When all the $B$-contours are straight lines, the field is termed quasisymmetric (QS). QS fields are sometimes classified into quasi-axisymmetric (QAS), quasi-poloidally symmetric (QPS) or quasi-helically symmetric fieds (QHS). In this paper, we would like to attempt a similar classification for pwO fields, to the extent to which this is possible. For instance, pwO fields do not have a specific helicity, as their $B_\mathrm{max}$-contours have a different topology.

This exercise has potential practical utility. In the case of omnigenous fields, within a given family (e.g. quasi-isodynamicity) different fields differ on the shape of the $B$-contours (except for $B=B_\mathrm{max}$). These $B$-contours can vary from one omnigenous field to another, although always in compliance with the constraints imposed by omnigenity. The so-called  
Cary \& Shasharina construction~\cite{cary1997omni}, discussed more in detail in~\cite{landreman2012omni}, provides a set of rules to construct acceptable $B$-contours. More recently, these rules have been parametrized, and included in optimization codes, and have led to the obtention of nearly omnigenous fields of all the families discussed in the previous paragraph \cite{dudt2023omni,liu2024omni}. In this work, we aim at parametrizing pwO fields in an analogous way. This is a first step towards including piecewise omnigenity as an explicit design criterion in stellarator optimization codes. 


The rest of the paper is organized as follows. We proceed from simpler to more complex and smoother (in terms of variation of $B$ on the flux surface) pwO fields. We start in section \ref{SEC_PWO} by revisiting the definition of piecewise omnigenity. We then characterize in detail, in section \ref{SEC_STRAIGHT}, the simplest possible pwO fields, already  presented in~\cite{velasco2024pwO}. For them, $B=B_\mathrm{max}$ inside a parallelogram and $B=B_0<B_\mathrm{max}$ outside it. These are pwO fields in which the $B$-contours consist of straight segments and $B$ is two-valued (and the minimum value of $B$ on the flux-surface is $B_\mathrm{min}=B_0$). As part of this characterization, in the different subsections of section \ref{SEC_STRAIGHT}, we discuss the constraints imposed by piecewise omnigenity (on the rotational transform, in particular) or by stellarator symmetry, and we compute the second adiabatic invariant for all trapped orbits. Then, in the next three sections, smoother pwO fields are presented by relaxing, alternatively, some of the unnecesary characteristics of the simple pwO fields of~\cite{velasco2024pwO}. First, in section \ref{SEC_NONSTRAIGHT}, we allow for the $B$-contours to be non-straight. Then, in sections \ref{SEC_HYBRIDS} and \ref{SEC_SCAN} we allow $B$ to take values in the range $[B_\mathrm{min}, B_0]$ outside the parallelogram in two different way: initially, in section \ref{SEC_HYBRIDS}, by \textit{combining} the simple pwO fields of \ref{SEC_STRAIGHT} with \textit{standard} omnigenous fields; then, in section \ref{SEC_SCAN}, by slightly modifying some of the simple fields of section \ref{SEC_STRAIGHT}. All the techniques discussed in the previous sections are then combined in section \ref{SEC_SMOOTHEST}. Finally, the conclusions and future work come in section \ref{SEC_SUMMARY}.


\section{Definition of piecewise omnigenity}\label{SEC_PWO}

A magnetic field $\mathbf{B}$ with nested toroidal surfaces can be expressed in terms of $s$, a flux-surface label, $\alpha$, a field line label, and $l$, the arc-length along the magnetic field line. In these coordinates, $\mathbf{B}$ can be written as
\begin{equation} 
\mathbf{B}=\partial_s\Psi(s)\nabla s\times\nabla\alpha\,,
\end{equation}
where $2\pi\Psi(s)$ is the toroidal flux enclosed by the flux surface $s$. Specifically, we will be employing $s=\Psi/\Psi_\mathrm{LCFS}$, where $\Psi_\mathrm{LCMS}=\Psi$ at the last closed flux-surface, and $\alpha=\theta-\iota\zeta$, where $\iota$ is the rotational transform. Note that, since we will be working in Boozer coordinates, field lines are straight lines of constant $\theta-\iota\zeta$.

Omnigenity is best discussed in terms of the second adiabatic invariant, defined for trapped particles as

\begin{equation}
J(s,\alpha,{\mathcal{E}},\mu) \equiv 2\int_{l_{b_1}}^{l_{b_2}}\mathrm{d}l \sqrt{2\left(\mathcal{E}-\mu B\right)} = 2\int_{\zeta_{b_1}}^{\zeta_{b_2}}\mathrm{d}\zeta \frac{I_p+\iota I_t}{B^2} \sqrt{2\left(\mathcal{E}-\mu B\right)}\,.\label{EQ_J}
\end{equation}
Here, ${\mathcal{E}}$ and $\mu$ are the energy and magnetic moment of the particle, respectively, $l_{b_1}$ and $l_{b_2}$ (and $\zeta_{b_1}$ and $\zeta_{b_2}$) label the bounce points of the trajectory (i.e., the points where the magnetic field strength $B$ is equal to ${\mathcal{E}}/\mu$ and the component of the velocity that is parallel to the field vanishes). Finally, $I_t(s)=\mathbf{B}\cdot\mathbf{e}_\theta$ and $I_p(s)=\mathbf{B}\cdot\mathbf{e}_\zeta$, where $\mathbf{e}_\theta$ and $\mathbf{e}_\theta$ are the covariant basis vectors in Boozer coordinates. The second adiabatic invariant characterizes the periodic trajectory of particles trapped in a stellarator field since, in the absence of collisions, particles move at constant $J$. Specifically, if $\mathbf{v}_d$ denotes the drift velocity perpendicular to $\mathbf{B}$, the following relations hold between the orbit-averaged radial and tangential drifts and the spatial variation of $J$:
\begin{eqnarray} 
\overline{\mathbf{v}_d\cdot\nabla s} = \frac{m}{\tau_b Ze\Psi_\mathrm{LCMS}}\partial_\alpha J\,,\quad\overline{\mathbf{v}_d\cdot\nabla \alpha} = -\frac{m}{\tau_b Ze\Psi_\mathrm{LCMS}}\partial_s J\,.
\end{eqnarray}
Here, $\overline{(...)}=(2/\tau_b)\int_{l_{b_1}}^{l_{b_2}}\mathrm{d}l (...)|v_\parallel|^{-1}$ denotes orbit-average, $\tau_b\equiv 2\int_{l_{b_1}}^{l_{b_2}}\mathrm{d}l/|v_\parallel|$ is the bounce time, and $Ze$ and $m$ are the particle charge and mass respectively. It is straightforward to see that, if $\partial_\alpha J=0$ for all trapped particles, the magnetic field is omnigenous. Equivalently, in an omnigenous field, $J$ is a flux function:
\begin{eqnarray}
J&=& J(s,{\mathcal{E}},\mu),\label{EQ_O}
\end{eqnarray}
(with the exception of the fields of \cite{parra2015omni}). In the presence of collisions, omnigenous fields display no radial energy flux associated to the $1/\nu$ regime \cite{nemov1999neo} as, in this stellarator-specific neoclassical regime, transport is proportional to $\partial_\alpha J$ and inversely proportional to the collision frequency. The so-called effective ripple is zero, as in an axisymmetric tokamak.

We call a field piecewise omnigenous if
\begin{eqnarray}
J&=& J^\mathrm{(w)}(s,{\mathcal{E}},\mu),\quad \mathrm{w}=\mathrm{I,II,III},\label{EQ_PWO}\\
\lim_{\Delta\alpha\to 0}J(s,\alpha+\Delta\alpha,{\mathcal{E}},\mu)&-&J(s,\alpha,{\mathcal{E}},\mu) \ne 0 \text{ at transitions between regions. }\label{EQ_TRANSITIONS}
\end{eqnarray}
In a pwO field, for a given particle velocity, there exist several classes of trapped particles, labelled by discrete index $\mathrm{w}$ in equation (\ref{EQ_PWO}). Within a region, $\partial_\alpha J=0$  for all particles. Trapped particles \textit{transition} between classes and encounter discontinuities in $J$, as indicated by equation (\ref{EQ_TRANSITIONS}). However, as discussed in~\cite{velasco2024pwO,calvo2024pwO}, when both equations are fulfilled, the field has zero effective ripple.

\begin{figure}
\includegraphics[angle=0,width=.45\columnwidth]{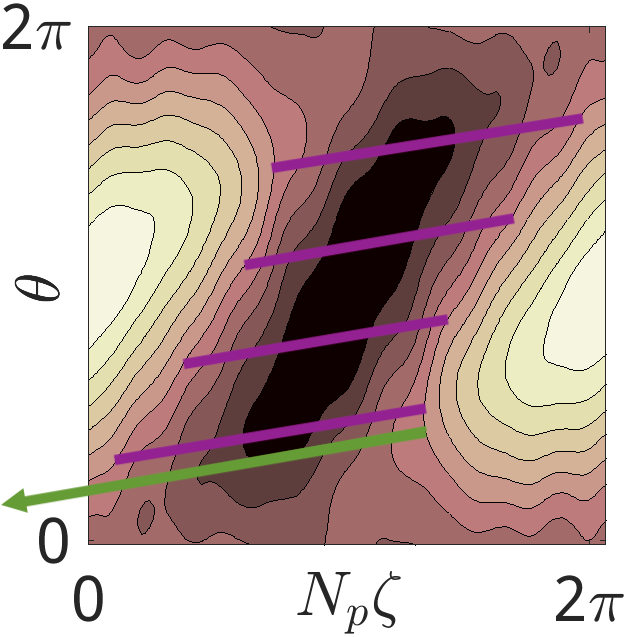}  
\includegraphics[angle=0,width=.45\columnwidth]{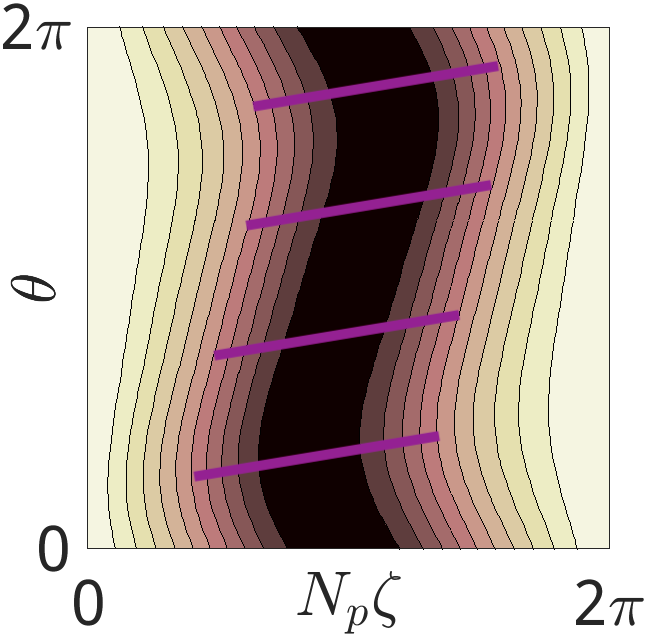}
\caption{Magnetic field strength on a flux-surface of W7-X standard (left) and a W7-X-like exactly QI field (right). Purple and green lines represent trapped particle orbits and a darker colour corresponds to a weaker field. \label{FIG_TRANSITION}}
\end{figure}

According to equation (\ref{EQ_J}), $J$ depends on $B(\theta,\zeta)$ through its integrand and, more crucially, through the position of the bounce points on the flux surface. It is through this dependence that equation (\ref{EQ_O}) affects the topology of the $B$-contours, $(\zeta_{b_1},\theta_{b_1})$ and $(\zeta_{b_2},\theta_{b_2})$: if they do not close toroidally, poloidally or helically, a particle that is bouncing back and forth along the field line and slowly drifting on the flux-surface will find one of the bounce points \textit{disappearing}. This is illustrated in figure \ref{FIG_TRANSITION} (left) by the green line, whose left bounce point is now in another field period (compare with figure \ref{FIG_TRANSITION} (right), with all particles trapped within one field period). These transitions cause transport in a generic stellarator (see e.g. \cite{dherbemont2022las}). In $J$, they cause discontinuities, which are forbidden by omnigenity, see equation (\ref{EQ_O}), but allowed by piecewise omnigenity, see equation (\ref{EQ_TRANSITIONS}). This leads to qualitatively different $B$-contours in a pwO field, as we will see in the next sections.

We end this section by noting that piecewise omnigenity can not be achieved with analytical fields. In practical terms, however, optimizing with respect to piecewise omnigenity reduces the $1/\nu$ flux, as ilustrated in~\cite{velasco2024pwO}. It remains to be studied what happens at lower collisionalities. For instance, the $\sqrt{\nu}$ regime has only been demonstrated to exist in stellarators close to omnigenity, but not in large-aspect ratio stellarators far from omnigenity \cite{dherbemont2022las}. In any case, exact piecewise omnigenity is to be considered an \textit{ideal} design goal.

\begin{figure}
\includegraphics[angle=0,width=\columnwidth]{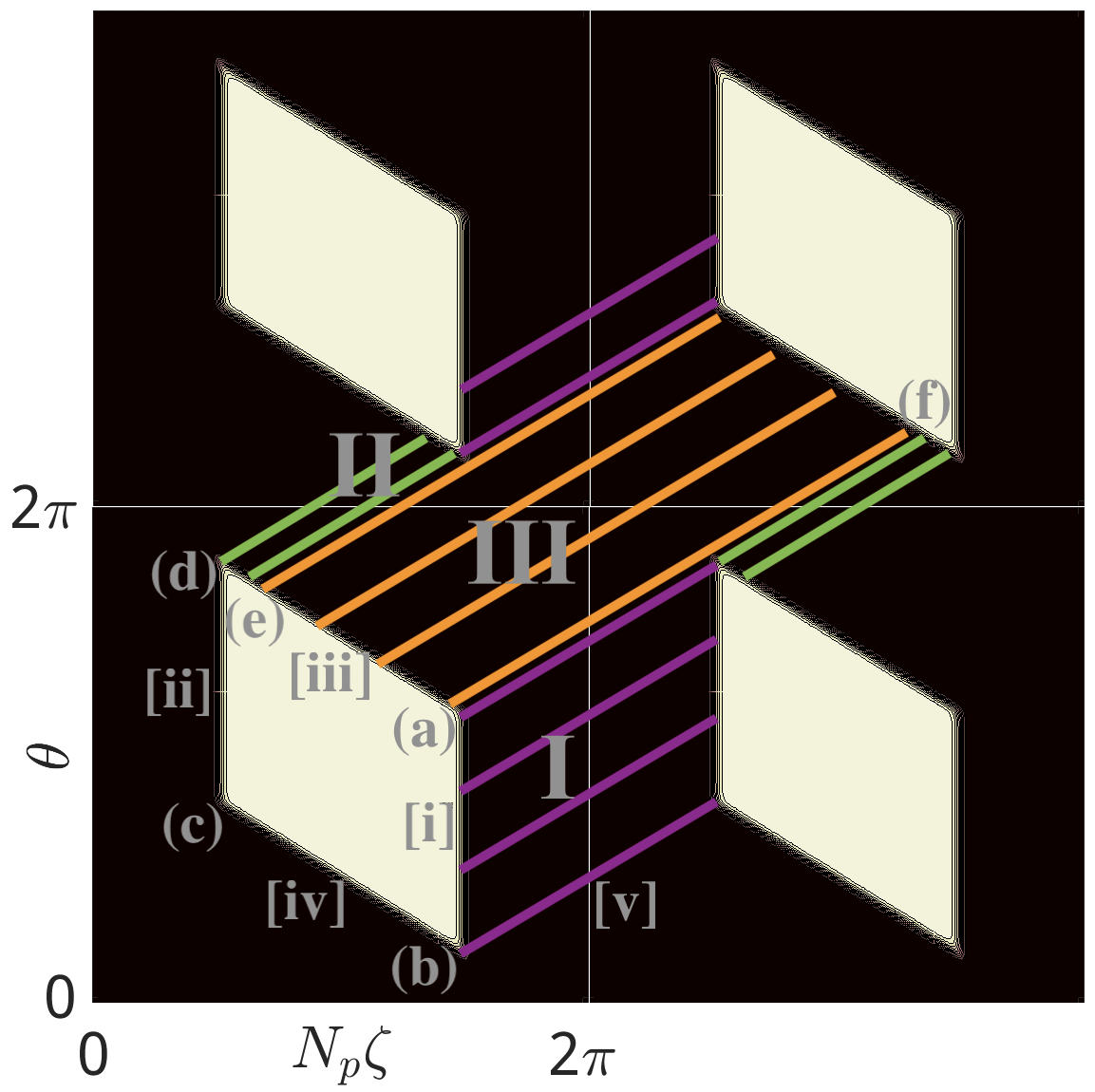}  
\caption{$B_{pwQS}$ of equations (\ref{EQ_BPWQS}) (with $p=15$ as an approximation to $p\to\infty$) and (\ref{EQ_IOTA0}) with $N_pw_1=w_2=\pi/2$, $t_1=0$, $t_2=\iota_0=\iota=1.242$, $\zeta_c=\pi/N_p$ and $\theta_c=\pi$. White corresponds to $B=B_\mathrm{max}$ and black to $B=B_\mathrm{min}$.~LEGEND:
(a) $(\zeta_{++},\theta_{++})$,
(b) $(\zeta_{+-},\theta_{+-})$,
(c) $(\zeta_{--},\theta_{--})$,
(d) $(\zeta_{-+},\theta_{-+})$,
(e) $(\zeta_{t+},\theta_{t+})$,
(f) $(\zeta_{t-}+2\pi/N_p,\theta_{t-}+2\pi)$,
[i] $\zeta-\zeta_c+t_1(\theta-\theta_c)=+w_1$,
[ii] $\zeta-\zeta_c+t_1(\theta-\theta_c)=-w_1$,
[iii] $\theta-\theta_c+t_2(\zeta-\zeta_c)=+w_2$,
[iv] $\theta-\theta_c+t_2(\zeta-\zeta_c)=-w_2$,
[v] $\theta-\theta_{+-}= \iota_0(\zeta-\zeta_{+-})$.
\label{FIG_PWQS}}
\end{figure}

\section{Simplest pwO fields: $B$-contours consisting of straight segments and $B=B_\mathrm{max}$ or $B=B_\mathrm{min}$}\label{SEC_STRAIGHT}

In~\cite{velasco2024pwO}, a set of sufficient (but not necessary, as we will confirm in sections \ref{SEC_NONSTRAIGHT} and  \ref{SEC_SCAN}) conditions for piecewise omnigenity was derived:
\begin{enumerate}
\item All the contours of constant $B$ must collapse into a single parallelogram.
\item The rotational transform has to be such that only two field lines connect the four corners (these corners may be located in different field periods).
\end{enumerate}

If conditions 1 and 2 are satisfied, as they are in the field of figure \ref{FIG_PWQS}, the magnetic field is piecewise omnigenous.

We construct our pwO field with $B$-contours consisting of straight segments, $B_{pwQS}$, as:
\begin{eqnarray}
B_{pwQS} &=&B_{0}+(B_\mathrm{max}-B_{0})\lim_{p\to\infty}  e^{-\left(\frac{\zeta-\zeta_c+t_1(\theta-\theta_c)}{w_1}\right)^{2p}-\left(\frac{\theta-\theta_c+t_2(\zeta-\zeta_c)}{w_2}\right)^{2p}}\,,\label{EQ_BPWQS}\\
\iota&=&\iota_0\equiv \left(\frac{\pi(1-t_1t_2)}{N_pw_1}-1\right)^{-1}t_2\,.\label{EQ_IOTA0}
\end{eqnarray}
Because its $B$-contours are consisting of straight segments, we label the field of equations (\ref{EQ_BPWQS}) and (\ref{EQ_IOTA0}) \textit{piecewise quasisymmetric}, (pwQS). The magnetic field of equation (\ref{EQ_BPWQS}) is defined in a toroidal field period, and continuity at the ends of the periods and periodicity are imposed afterwards. An example, with a specific choice of parameters (see caption), is depicted in figure \ref{FIG_PWQS}. The role of the different parameters will become apparent in the following subsections. In them, we will confirm that equations (\ref{EQ_BPWQS}) and (\ref{EQ_IOTA0}), and thus the example of figure \ref{FIG_PWQS}, achieve the conditions of equations (\ref{EQ_PWO}) and (\ref{EQ_TRANSITIONS}).

\subsection{Shape of the $B$-contours}\label{SEC_SHAPE}

It is straightforward to see that, because
\begin{equation}
\lim_{p\to\infty}  e^{{-x}^{2p}}=
\begin{cases}
1 & \text{if } x < 1, \\
0 & \text{if } x > 1,
\end{cases}
\end{equation}
all the $B_\mathrm{min}=B_0 < B < B_\mathrm{max}$ contours lie at angular positions where one of the following relations are fulfilled:
\begin{eqnarray}
\zeta-\zeta_c+t_1(\theta-\theta_c)=+w_1,\label{EQ_SIDEP1}\\
\zeta-\zeta_c+t_1(\theta-\theta_c)=-w_1,\label{EQ_SIDEM1}\\
\theta-\theta_c+t_2(\zeta-\zeta_c)=+w_2,\label{EQ_SIDEP2}\\
\theta-\theta_c+t_2(\zeta-\zeta_c)=-w_2.\label{EQ_SIDEM2}
\end{eqnarray}
Equations (\ref{EQ_SIDEP1}) and (\ref{EQ_SIDEM1}) correspond to parallel lines, with a tilting given by $-1/t_1$, toroidally separated a distance $2w_1$. Equations (\ref{EQ_SIDEP2}) and (\ref{EQ_SIDEM2}) correspond to parallel lines, with a tilting given by $-t_2$, poloidally separated a distance $2w_2$. These four lines interesect at four corners that define a parallelogram. Condition 1 for piecewise omnigenity is thus fulfilled.


\subsection{Constraints imposed by the rotational transform}\label{SEC_IOTA}

The four corners of the parallelogram can be obtained by combining equations (\ref{EQ_SIDEP1}) and (\ref{EQ_SIDEP2}),
\begin{eqnarray}
\zeta_{++}&=& \zeta_c+\frac{w_1 -t_1 (\theta_c+w_2)}{1-t_1t_2},\\
\theta_{++}&=& \theta_c+\frac{w_2 -t_2 (\zeta_c+w_1)}{1-t_1t_2},
\end{eqnarray}
equations (\ref{EQ_SIDEP1}) and (\ref{EQ_SIDEM2}),
\begin{eqnarray}
\zeta_{+-}&=& \zeta_c+\frac{w_1 -t_1 (\theta_c-w_2)}{1-t_1t_2},\\
\theta_{+-}&=& \theta_c+\frac{-w_2 -t_2 (\zeta_c+w_1)}{1-t_1t_2},
\end{eqnarray}
equations (\ref{EQ_SIDEM1}) and (\ref{EQ_SIDEM2}),
\begin{eqnarray}
\zeta_{--}&=& \zeta_c+\frac{-w_1 -t_1 (\theta_c-w_2)}{1-t_1t_2},\\
\theta_{--}&=& \theta_c+\frac{-w_2 -t_2 (\zeta_c-w_1)}{1-t_1t_2},
\end{eqnarray}
and equations (\ref{EQ_SIDEM1}) and (\ref{EQ_SIDEP2}),
\begin{eqnarray}
\zeta_{-+}&=& \zeta_c+\frac{-w_1 -t_1 (\theta_c+w_2)}{1-t_1t_2},\\
\theta_{-+}&=& \theta_c+\frac{+w_2 -t_2 (\zeta_c-w_1)}{1-t_1t_2}.
\end{eqnarray}

The corners $(\zeta_{++},\theta_{++})$ and $(\zeta_{-+}+2\pi/N_p,\theta_{-+})$ are connected by a field line if
\begin{eqnarray}
  \iota=\frac{\theta_{-+}-\theta_{++}}{\zeta_{-+}+2\pi/N_p-\zeta_{++}}=\left(\frac{\pi(1-t_1t_2)}{N_pw_1}-1\right)^{-1}t_2=\iota_0.
\end{eqnarray}
Because a parallelogram has opposing sides of equal length, once $(\zeta_{++},\theta_{++})$ and $(\zeta_{-+}+2\pi/N_p,\theta_{-+})$ are connected by a field line, $(\zeta_{+-},\theta_{+-})$ and $(\zeta_{--}+2\pi/N_p,\theta_{--})$ are automatically connected by another field line. Condition 2 for piecewise omnigenity is thus fulfilled.

We note that the relation
\begin{eqnarray}
\iota=\iota_0(t_1,t_2,N_pw_1)
\end{eqnarray}
can be interpreted either as a constraint that the parameters of equation (\ref{EQ_BPWQS}) must fulfill for a given rotational transform $\iota$, or as the unique (in general, although see section \ref{SEC_ALTIOTA}) value of the rotational transform that achieves piecewise omnigenity once a field $B(\theta,\zeta)$ given by equation (\ref{EQ_BPWQS}) is specified (although values of $\iota$ close to $\iota_0$ can also result in reduced neoclassical transport, see \cite{velasco2024pwO}).

The result of fulfilling $\iota=\iota_0$ is that three classes of orbits exist on the flux-surface. Purple orbits (region I) have their bounce points on the segments given by equations (\ref{EQ_SIDEP1}) and (\ref{EQ_SIDEM1}). The segments given by equations (\ref{EQ_SIDEP2}) and (\ref{EQ_SIDEM2}) generally contain bounce points of two classes of orbits, green (region II) and orange (region III). Two points, $(\zeta_{t+},\theta_{t+})$ and $(\zeta_{t-},\theta_{t-})$, each in one segment, separate regions II and III. The former, for the case of figure \ref{FIG_PWQS}, is found by intersecting the segment described by equation (\ref{EQ_SIDEP2}) and a field line in which the corner $(\zeta_{+-},\theta_{+-}+2\pi)$ lies,
\begin{eqnarray}
\theta-\theta_{+-}-2\pi=\iota_0 (\zeta-\zeta_{+-});
\end{eqnarray}
the latter is found by intersecting the segment described by equation (\ref{EQ_SIDEM2}) and a field line that contains the corner $(\zeta_{-+},\theta_{-+}-2\pi)$,
\begin{eqnarray}
\theta-\theta_{-+}+2\pi=\iota_0 (\zeta-\zeta_{-+}).
\end{eqnarray}
The result is
\begin{eqnarray}
\theta_{t+}&=&\frac{\iota_0}{\iota_0+t_2}\left(\theta_c+t_2\zeta_c+w_2-\frac{t_2}{\iota_0}(\iota_0\zeta_{+-}-\theta_{+-}-2\pi)\right),\\
\zeta_{t+}&=&\frac{\theta_c+t_2\zeta_c+w_2-\theta_{t+}}{t_2},\\
\theta_{t-}&=&\frac{\iota_0}{\iota_0+t_2}\left(\theta_c+t_2\zeta_c-w_2-\frac{t_2}{\iota_0}(\iota_0\zeta_{-+}-\theta_{-+}+2\pi)\right),\\
\zeta_{t-}&=&\frac{\theta_c+t_2\zeta_c-w_2-\theta_{t-}}{t_2}.
\end{eqnarray}
These particular expressions are valid for the case in which particles in region II or III have their bounce points less than one field period away. In other situations (see section \ref{SEC_ALTIOTA}), $(\zeta_{t+},\theta_{t+})$ and $(\zeta_{t-},\theta_{t-})$ have to be found by intersecting equations (\ref{EQ_SIDEP2}) and (\ref{EQ_SIDEM2}) with field lines that contain corners of different field periods.

It is useful to define $f_\mathrm{II}=f_\mathrm{II}(\zeta_c,\theta_c,t_1,t_2,w_1,w_2,N_p)$ as the fraction of the segment that corresponds to region II:
\begin{eqnarray}
f_\mathrm{II}=\frac{\zeta_{t+}-\zeta_{-+}}{\zeta_{++}-\zeta_{-+}}=\frac{\zeta_{+-}-\zeta_{t-}}{\zeta_{+-}-\zeta_{--}}.
\end{eqnarray}
The rest, $1-f_\mathrm{II}$, corresponds to region III.

\subsection{Alternative values of the rotational transform}\label{SEC_ALTIOTA}

There are alternative ways of shaping $B(\zeta,\theta)$, while still fulfilling equations (\ref{EQ_PWO}) and (\ref{EQ_TRANSITIONS}), to those discussed in section \ref{SEC_IOTA}. For some specific choices of the parameters for equation (\ref{EQ_PWO}), there are alternative values of the rotational transform that are consistent with piecewise omnigenity. For instance, instead of connecting the corners $(\zeta_{++},\theta_{++})$ and $(\zeta_{-+}+2\pi/N_p,\theta_{-+})$, one could make field lines connect $(\zeta_{++},\theta_{++})$ and $(\zeta_{+-},\theta_{+-}+2\pi)$. This gives
\begin{eqnarray}
\iota=\iota_1\equiv \frac{\pi(1-t_1t_2)-w_2}{t_1w_2}.
\end{eqnarray}
This is typically a large value of $\iota/N_p$, and it is perhaps only feasable for $N_p=1$. Slightly larger values of $N_p$ may be possible if the field lines are made connect the corners $(\zeta_{++},\theta_{++})$ and $(\zeta_{+-}+2\pi/N_p,\theta_{+-}+2\pi)$. This option, only feasable if $w_1$ and/or $w_2$ are small enough, gives
\begin{eqnarray}
\iota=\iota_2\equiv \frac{\pi(1-t_1t_2)-w_2}{t_1w_2+\pi(1-t_1t_2)/N_p}.
\end{eqnarray}

We will not consider these or other alternative cases in the rest of this manuscript. We will also not consider $B(\theta,\zeta)$ consisting on a region with $B=B_\mathrm{min}$ and more than one disconnected regions with $B=B_\mathrm{max}$~\cite{escoto2024pwO}.

Finally, there exists the case, mentioned in~\cite{velasco2024pwO}, in which two of the sides of the parallelogram are parallel to the field lines. This happens when
\begin{eqnarray}
\iota=\iota_3\equiv -\frac{1}{t_1},
\end{eqnarray}
or
\begin{eqnarray}
\iota=\iota_4\equiv -t_2.
\end{eqnarray}
When this is the case, $J$ is piecewisely constant on the flux-surface, and there exist transitioning particles. However, the alignment of the field lines with the sides of the parallelogram produces non-zero $1/\nu$ transport~\cite{calvo2024pwO}, and thus these fields will not be included in this study devoted to piecewise omnigenous fields.

\subsection{Stellarator symmetry}\label{SEC_SSYMMETRY}

Stellarator symmetry requires the center of the parallelogram to lie at $\zeta_c=k_1\pi/N_p$, $\theta_c=k_2\pi$, with $k_i=0,1$. It is straightforward to check that this guarantees that, under the operation $(\zeta,\theta)\rightarrow (2\pi/N_p-\zeta,2\pi-\theta)$, the segment given by equation (\ref{EQ_SIDEP1}) transforms into the segment given by equation (\ref{EQ_SIDEM1}) and viceversa. Similarly, the segment given by equation (\ref{EQ_SIDEP2}) transforms into the segment given by equation (\ref{EQ_SIDEM2}) and viceversa. As a consequence of this, corners $(\zeta_{++},\theta_{++})$ and $(\zeta_{+-},\theta_{+-})$ transform into $(\zeta_{--},\theta_{--})$ and $(\zeta_{-+},\theta_{-+})$ and viceversa.

\subsection{Second adiabatic invariant}\label{SEC_J}

The expression defining the second adiabatic invariant (see equation (\ref{EQ_J})) can be computed explicitly for the magnetic field of equations (\ref{EQ_BPWQS}) and (\ref{EQ_IOTA0}). It essentially reduces to computing the distance along the field line between the trajectory bounce points,
\begin{eqnarray}
\Delta
\zeta_b^\mathrm{(w)}\equiv \zeta_{b_2}^\mathrm{(w)}-\zeta_{b_1}^\mathrm{(w)}.\label{EQ_JW}
\end{eqnarray}
Then,
\begin{eqnarray}
  J^\mathrm{(w)}=2\Delta\zeta_b^\mathrm{(w)}  \frac{I_p+\iota I_t}{B_{0}^2} \sqrt{2\left(\mathcal{E}-\mu B_{0}\right)}\,.
\end{eqnarray}
For region I (purple orbits), we have
\begin{eqnarray}
  \Delta\zeta_b^\mathrm{(I)}=\zeta_{-+}+\frac{2\pi}{N_p}-\zeta_{++}=\frac{2\pi}{N_p}-\frac{2w_1}{1-t_1t_2}=\frac{2\pi}{N_p}\left(1-\frac{N_pw_1}{\pi(1-t_1t_2)}\right).\label{EQ_JI}
\end{eqnarray}
In order to compute $\Delta\zeta_b^\mathrm{(II)}$, we start by noting that the bounce points on region II (green) lie on the segments described by equation (\ref{EQ_SIDEP2}) and
\begin{eqnarray}
\theta-\theta_c-2\pi+t_2(\zeta-\zeta_c)=-w_2.
\end{eqnarray}
These segments are  at a poloidal distance $2(\pi-w_2)$, and the shortest segment that connects them has constant $t_2\theta-\zeta$. Using that this line forms a $\tan^{-1}(t_2)$ angle with the vertical line and a $\tan^{-1}[(t_2\iota-1)/(t_2+\iota)]$ angle with the field lines, and that $\cos(\tan^{-1}(x))=(1+x^2)^{-1/2}$, it can be obtained that
\begin{eqnarray}
  \Delta \zeta_b^\mathrm{(II)} &=& 2(\pi-w_2)\frac{\cos[\tan^{-1}(t_2)]\cos[\tan^{-1}(\iota)]}{\cos[\tan^{-1}((t_2\iota-1)/(t_2+\iota))]}=\frac{2(\pi-w_2)}{t_2+\iota},\label{EQ_JII}
\end{eqnarray}
which, using equation (\ref{EQ_IOTA0}), can be rewritten as
\begin{eqnarray}
  \Delta \zeta_b^\mathrm{(II)} =\frac{2(\pi-w_2)}{t_2}\left(1-\frac{N_pw_1}{\pi(1-t_1t_2)}\right).\label{EQ_JIIb}
\end{eqnarray}

Similarly, for region III (orange orbits), the bounce points lie on the segments described by equation (\ref{EQ_SIDEP2}) and
\begin{eqnarray}
\theta-\theta_c-2\pi+t_2(\zeta-\zeta_c-2\pi/N_p)=-w_2.
\end{eqnarray}
and are at a poloidal distance  $2[\pi(1+t_2/N_p)-w_2]$. Calculations analogous to those of region II yield
\begin{eqnarray}
  \Delta \zeta_b^\mathrm{(III)} =\frac{2[\pi(1+t_2/N_p)-w_2]}{t_2}\left(1-\frac{N_pw_1}{\pi(1-t_1t_2)}\right).\label{EQ_JIII}
\end{eqnarray}
Comparison of equations (\ref{EQ_JI}), (\ref{EQ_JII}) and (\ref{EQ_JIII}) show that 
\begin{eqnarray}
\Delta \zeta_b^\mathrm{(III)} &=&  \Delta\zeta_b^\mathrm{(I)} + \Delta\zeta_b^\mathrm{(II)},\label{EQ_JCHECK}
\end{eqnarray}
as expected.

It is straightforward to take the radial derivative of equation (\ref{EQ_JW}) for each region w, in order to study its dependence on parameters such as $\zeta_c$, $\theta_c$, $t_i$, or $w_i$. Quantity $\partial_s J$ is relevant for the study of fast ion confinement, see e.g.~\cite{velasco2021prompt}, which is left for a future work.

\section{pwO fields with $B$-contours consisting of non straight segments}\label{SEC_NONSTRAIGHT}

\begin{figure}
\includegraphics[angle=0,width=0.49\columnwidth]{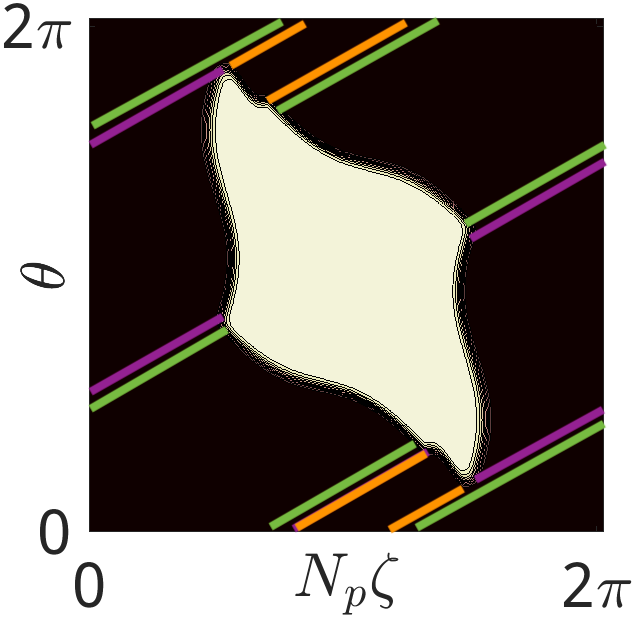}  
\includegraphics[angle=0,width=0.49\columnwidth]{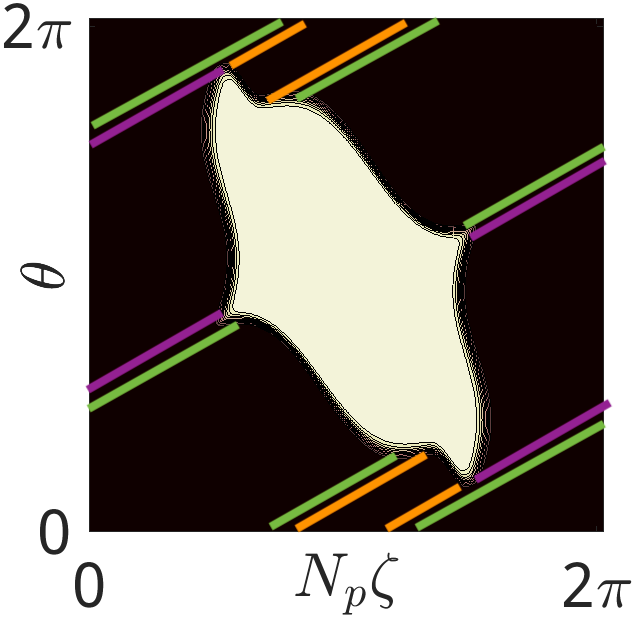}
\includegraphics[angle=0,width=0.49\columnwidth]{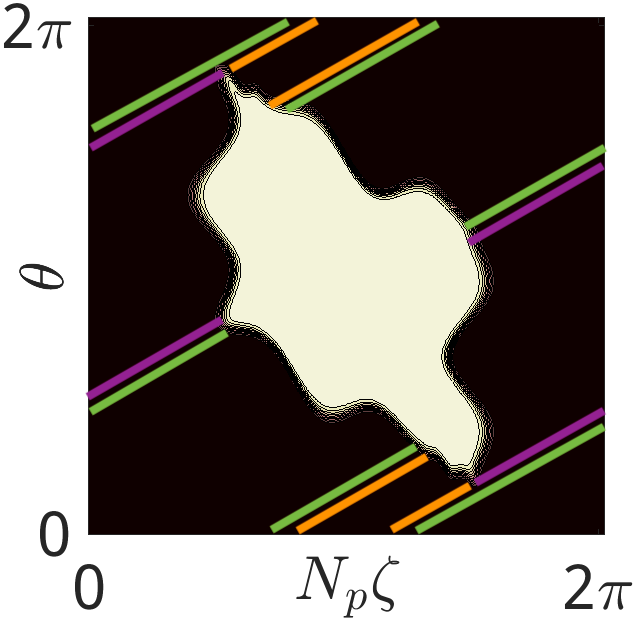}
\includegraphics[angle=0,width=0.49\columnwidth]{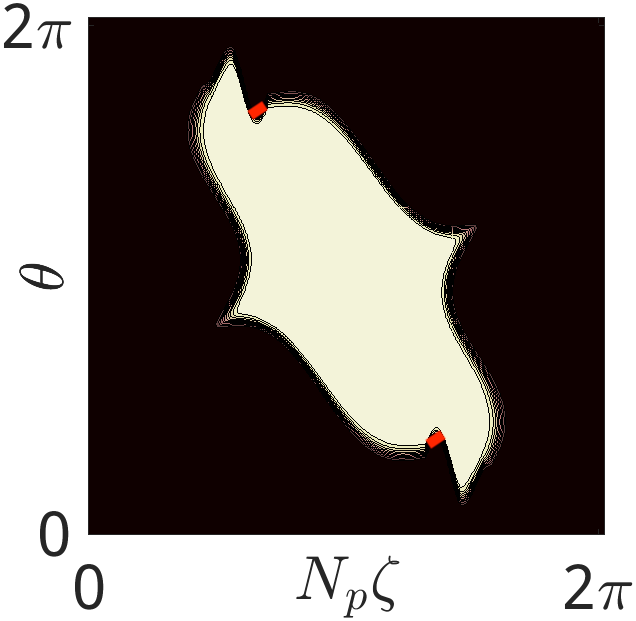}
\caption{$B_{pwO}$ of equation (\ref{EQ_BPWO}) with same parameters of figure \ref{FIG_PWQS} and, additionally, $s_1^\mathrm{(I)}=0.1w_1$, $s_1^\mathrm{(II)}=0.05w_2$, $s_1^\mathrm{(III)}=-0.1w_2$ (top left), $s_1^\mathrm{(I)}=0.1w_1$, $s_1^\mathrm{(II)}=0.05w_2$, $s_1^\mathrm{(III)}=s_1^\mathrm{(II)}(1-f_\mathrm{II})/f_\mathrm{II}$ (top right), $c_1^\mathrm{(I)}=-c_2^\mathrm{(I)}=0.1w_1$, $c_1^\mathrm{(II)}=-c_2^\mathrm{(II)}=0.02w_2$, $c_1^\mathrm{(III)}=-c_2^\mathrm{(III)}=0.1w_2$ (bottom left), $s_1^\mathrm{(I)}=0.2w_1$, $s_1^\mathrm{(II)}=0.2w_2$, $s_1^\mathrm{(III)}=0.2w_2$ (bottom right).\label{FIG_PWO}}
\end{figure}

The fields of section \ref{SEC_STRAIGHT}, which are consisting of straight segments, can be generalised to have non-straight segments. We would like to find such a magnetic field that, combined with equation (\ref{EQ_IOTA0}), still fulfills equations (\ref{EQ_PWO}) and (\ref{EQ_TRANSITIONS}). Specifically, we are interested in pwO fields whose contours of constant $B$ are not consisting of straight segments but have the same corners as equation (\ref{EQ_BPWQS}) (this may be done without loss of generality, as the corners may be modified through $\zeta_c$, $\theta_c$, $w_i$, and $t_i$, and $\iota$ would need to change accordingly, as described in section~\ref{SEC_STRAIGHT}). We construct our pwO fields (that are shown in figure (\ref{FIG_PWO})) as
\begin{eqnarray}
  B_{pwO} &=&B_{0}+(B_\mathrm{max}-B_{0})\lim_{p\to\infty} e^{-\left(\frac{\zeta-\zeta_c+t_1(\theta-\theta_c)}{w_1+\delta w_1(\zeta,\theta)}\right)^{2p}-\left(\frac{\theta-\theta_c+t_2(\zeta-\zeta_c)}{w_2+\delta w_2(\zeta,\theta)}\right)^{2p}}\,,\label{EQ_BPWO}
\end{eqnarray}
with
\begin{eqnarray}
\delta w_1(\zeta,\theta)&=& \sum_{n\ge 0}\left[s_{n}^\mathrm{(I)}\sin\left(2\pi n\frac{\theta-\theta_{+-}}{\theta_{++}-\theta_{+-}}\right)+\right.\nonumber\\
  & & ~~~ \left. + c_{n}^\mathrm{(I)}\cos\left(2\pi n\frac{\theta-\theta_{+-}}{\theta_{++}-\theta_{+-}}\right)\right] ~~~\text{if } \zeta-\zeta_c+t_1(\theta-\theta_c)>0,\label{EQ_SIDEPI}\\
\delta w_1(\zeta,\theta)&=& -\sum_{n\ge 0}\left[s_{n}^\mathrm{(I)}\sin\left(2\pi n\frac{\theta-\theta_{--}}{\theta_{-+}-\theta_{--}}\right)+\right.\nonumber\\
  & & ~~~ \left. + c_{n}^\mathrm{(I)}\cos\left(2\pi n\frac{\theta-\theta_{--}}{\theta_{-+}-\theta_{--}}\right)\right] ~~~\text{if } \zeta-\zeta_c+t_1(\theta-\theta_c)<0,\label{EQ_SIDEMI}\\
\delta w_2(\zeta,\theta)&=& \sum_{n\ge 0}\left[s_{n}^\mathrm{(II)}\sin\left(2\pi n\frac{\zeta-\zeta_{-+}}{\zeta_{t+}-\zeta_{-+}}\right)+\right.\nonumber\\
  & & ~~~ \left. + c_{n}^\mathrm{(II)}\cos\left(2\pi n\frac{\zeta-\zeta_{-+}}{\zeta_{t+}-\zeta_{-+}}\right)\right] ~~~\text{if } \theta-\theta_c+t_2(\zeta-\zeta_c)>0,~~\zeta<\zeta_{t+},\label{EQ_SIDEPII}\\
\delta w_2(\zeta,\theta)&=& \sum_{n\ge 0}\left[s_{n}^\mathrm{(III)}\sin\left(2\pi n\frac{\zeta-\zeta_{t+}}{\zeta_{++}-\zeta_{t+}}\right)+\right.\nonumber\\
  & & ~~~ \left. + c_{n}^\mathrm{(III)}\cos\left(2\pi n\frac{\zeta-\zeta_{t+}}{\zeta_{++}-\zeta_{t+}}\right)\right] ~~~\text{if } \theta-\theta_c+t_2(\zeta-\zeta_c)>0,~~\zeta>\zeta_{t+},\label{EQ_SIDEPIII}\\
\delta w_2(\zeta,\theta)&=& -\sum_{n\ge 0}\left[s_{n}^\mathrm{(II)}\sin\left(2\pi n\frac{\zeta-\zeta_{t-}}{\zeta_{+-}-\zeta_{t-}}\right)+\right.\nonumber\\
  & & ~~~~~~ \left. + c_{n}^\mathrm{(II)}\cos\left(2\pi n\frac{\zeta-\zeta_{t-}}{\zeta_{+-}-\zeta_{t-}}\right)\right] ~~~\text{if } \theta-\theta_c+t_2(\zeta-\zeta_c)<0,~~\zeta>\zeta_{t-},\label{EQ_SIDEMII}\\
\delta w_2(\zeta,\theta)&=& -\sum_{n\ge 0}\left[s_{n}^\mathrm{(III)}\sin\left(2\pi n\frac{\zeta-\zeta_{--}}{\zeta_{t-}-\zeta_{--}}\right)+\right.\nonumber\\
  & & ~~~~~~  \left. + c_{n}^\mathrm{(III)}\cos\left(2\pi n\frac{\zeta-\zeta_{--}}{\zeta_{t-}-\zeta_{--}}\right)\right] ~~~\text{if } \theta-\theta_c+t_2(\zeta-\zeta_c)<0,~~\zeta<\zeta_{t-},\label{EQ_SIDEMIII}
\end{eqnarray}
and
\begin{eqnarray}
\sum_{n\ge 0} (-1)^n c_{n}^\mathrm{(w)} = 0.\label{EQ_FIXCORNERS}
\end{eqnarray}
The  values of $s_n^\mathrm{w}$ and $c_n^\mathrm{w}$ are further constrained by
\begin{eqnarray}
\frac{\partial\delta w_1}{\partial\theta}&<&t_1+\frac{1}{\iota}\label{EQ_DW1}\\
\frac{\partial\delta w_2}{\partial\zeta}&<&\iota+t_2\label{EQ_DW2}
\end{eqnarray}
for $\iota>0$ and 
\begin{eqnarray}
\frac{\partial\delta w_1}{\partial\theta}&>&t_1+\frac{1}{\iota}\label{EQ_DW1b}\\
\frac{\partial\delta w_2}{\partial\zeta}&>&\iota+t_2\label{EQ_DW2b}
\end{eqnarray}
for $\iota<0$. Optionally, 
\begin{eqnarray}
\frac{\sum_{n>0} n s_{n}^\mathrm{(II)}}{f_\mathrm{II}}=\frac{\sum_{n>0} n s_{n}^\mathrm{(III)}}{1-f_\mathrm{II}}
\end{eqnarray}
ensures continuous derivatives of the $B$-contours at $(\zeta_{t+},\theta_{t+},)$ and $(\zeta_{t-},\theta_{t-},)$, and
\begin{eqnarray}
c_{n}^\mathrm{(w)}=0
\end{eqnarray}
is required by stellarator symmetry.

Equation (\ref{EQ_SIDEPI}) modifies the segment given by equation (\ref{EQ_SIDEP1}) in the most general way. Additional equation (\ref{EQ_FIXCORNERS}) guarantees that this is done in a way that the corner points do not move (this will also ensure continuity of the $B$-contours at $(\zeta_{t+},\theta_{t+},)$ and $(\zeta_{t-},\theta_{t-},)$). Equation (\ref{EQ_SIDEMI}) then modifies the segment given by equation (\ref{EQ_SIDEM1}) in a way that the distance along the field lines between bounce points at the curves given by equations (\ref{EQ_SIDEPI}) and (\ref{EQ_SIDEMI}) equals the distance along the field lines between bounce points at segments given by equations (\ref{EQ_SIDEP1}) and (\ref{EQ_SIDEM1}). This is guaranteed by
\begin{eqnarray}
  \delta w_1(\zeta_{b_1}^\mathrm{(I)},\theta_{b_1}^\mathrm{(I)})+\delta w_1(\zeta_{b_2}^\mathrm{(I)},\theta_{b_2}^\mathrm{(I)})=0,
\end{eqnarray}
due to
\begin{eqnarray}
\frac{\theta_{b_1}^\mathrm{(I)}-\theta_{+-}}{\theta_{++}-\theta_{+-}}=\frac{\theta_{b_2}^\mathrm{(I)}-\theta_{--}}{\theta_{-+}-\theta_{--}}.
\end{eqnarray}
Because each pair of bounce points, $(\zeta_{b_1}^\mathrm{(I)},\theta_{b_1}^\mathrm{(I)})$ and $(\zeta_{b_2}^\mathrm{(I)},\theta_{b_2}^\mathrm{(I)})$, is moved a same amount in the same direction, they end up connected by a (generally different) field line with the same separation. Finally, equations (\ref{EQ_DW1}) or (\ref{EQ_DW1b}) make sure that the field lines do not become tangent to the modified $B$-contours, which would destroy omnigenity  (a similar problem arises when trying to design omnigenous fields with toroidal symmetry too far from quasisymmetry, see the discussion at the end of section IV of \cite{landreman2012omni}). This happens when 
\begin{eqnarray}
\frac{1}{\iota}=-t_1+\frac{\partial\delta w_1}{\partial\theta}\,.
\end{eqnarray}
(in deriving  equations (\ref{EQ_DW1}) or (\ref{EQ_DW1b}), it has been assumed that $|t_1|\ll |\iota|$).

The segment of equation (\ref{EQ_SIDEP2}) is less straightforward to modify, as two different classes of trapped particles have their bounce points on it. The results of this modification are equations (\ref{EQ_SIDEPII}) and (\ref{EQ_SIDEPIII}). Then, equations (\ref{EQ_SIDEMII}) and (\ref{EQ_SIDEMIII}) deform the segment of equation (\ref{EQ_SIDEM2}) maintaining the distance between bounce points constant. Specifically,
\begin{eqnarray}
  \delta w_2(\zeta_{b_1}^\mathrm{(II)},\theta_{b_1}^\mathrm{(II)})+\delta w_2(\zeta_{b_2}^\mathrm{(II)},\theta_{b_2}^\mathrm{(II)})=0
\end{eqnarray}
and
\begin{eqnarray}
  \delta w_2(\zeta_{b_1}^\mathrm{(III)},\theta_{b_1}^\mathrm{(III)})+\delta w_2(\zeta_{b_2}^\mathrm{(III)},\theta_{b_2}^\mathrm{(III)})=0
\end{eqnarray}
follow from
\begin{eqnarray}
\frac{\zeta_{b_1}^\mathrm{(II)}-\zeta_{-+}}{\zeta_{t+}-\zeta_{-+}}=\frac{\zeta_{b_2}^\mathrm{(II)}-\zeta_{t-}}{\zeta_{+-}-\zeta_{t-}}
\end{eqnarray}
and
\begin{eqnarray}
  \frac{\zeta_{b_1}^\mathrm{(III)}-\zeta_{t+}}{\zeta_{++}-\zeta_{t+}}=\frac{\zeta_{b_2}^\mathrm{(III)}-\zeta_{--}}{\zeta_{t-}-\zeta_{--}}
\end{eqnarray}
respectively.  Equations (\ref{EQ_DW2}) and (\ref{EQ_DW2b}) make sure that the field lines do not become tangent to the modified $B$-contours, which happens when 
\begin{eqnarray}
{\iota}=-t_2+\frac{\partial\delta w_2}{\partial\zeta}\,.
\end{eqnarray}
(in deriving  equations (\ref{EQ_DW2}) and (\ref{EQ_DW2b}), it has been assumed that $t_2\sim\iota$ and $t_2\iota>0$).

Examples of fields for equation (\ref{EQ_BPWO}) are shown in figure \ref{FIG_PWO}. Figure \ref{FIG_PWO} (top left) shows a stellarator-symmetric example with discontinuous derivatives of the $B$-contours at $(\zeta_{t+},\theta_{t+},)$ and $(\zeta_{t-},\theta_{t-},)$. Figure \ref{FIG_PWO} (top right) shows a stellarator-symmetric example with continuous derivatives of the $B$-contours at $(\zeta_{t+},\theta_{t+},)$ and $(\zeta_{t-},\theta_{t-},)$. Figure \ref{FIG_PWO} (bottom left) shows an example without stellarator symmetry. Finally, \ref{FIG_PWO} (bottom right) shows an example in which the $B$-contours deviate too much from a parallelogram, equation (\ref{EQ_DW2}) is not fulfilled, and non-omnigenous small wells are created (marked in red).

\begin{figure}
\includegraphics[angle=0,width=0.49\columnwidth]{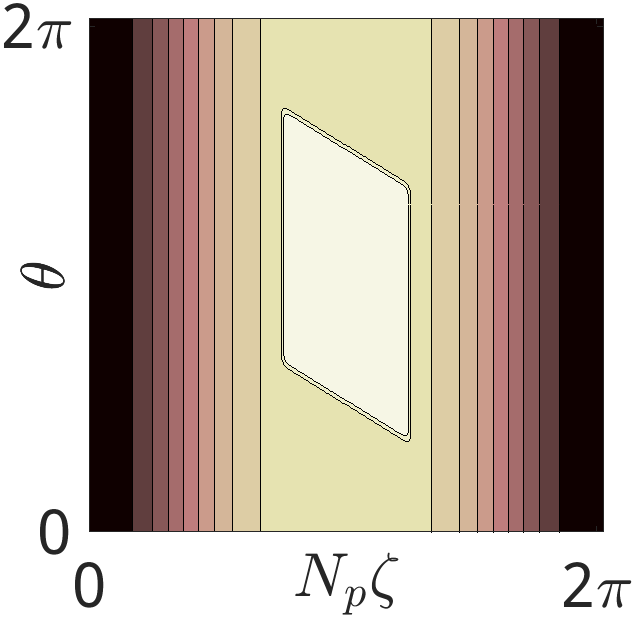}  
\includegraphics[angle=0,width=0.49\columnwidth]{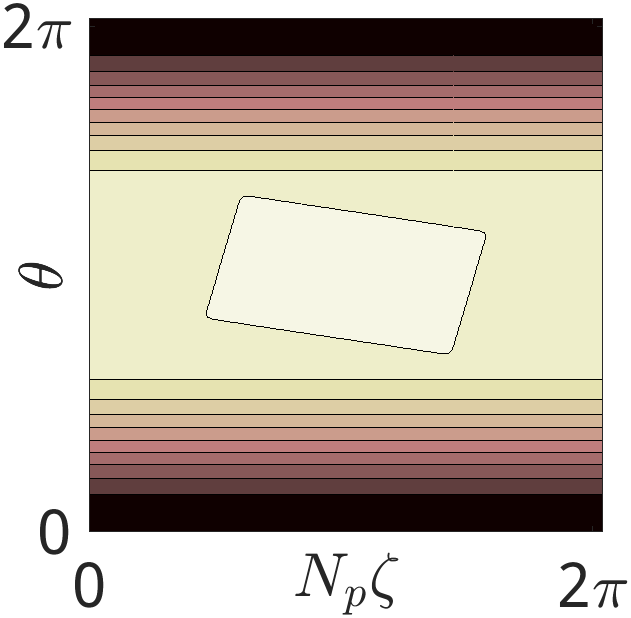}  
\includegraphics[angle=0,width=0.49\columnwidth]{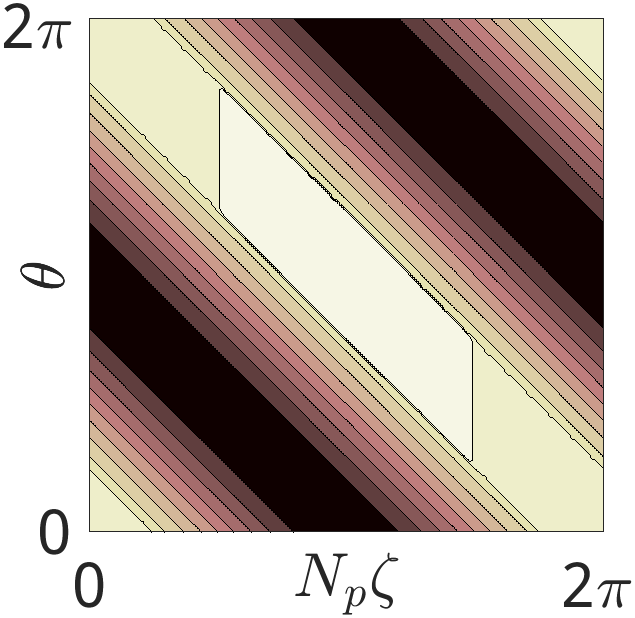}  
\includegraphics[angle=0,width=0.49\columnwidth]{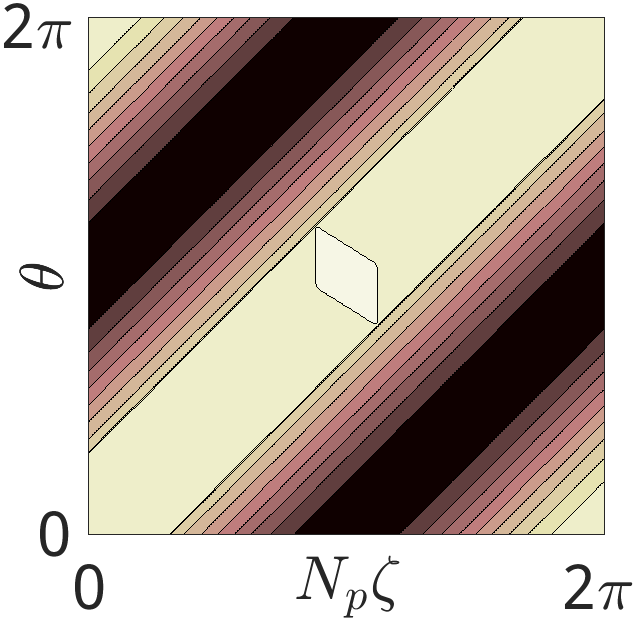}  
\caption{Examples of fields that are partially quasisymmetric and partially piecewise quasisymmetric: $M=0$ (top left), $N=0$ (top right), $M=1$, $N=-1$ (bottom left) and $M=1$, $N=1$ (bottom right). White represents $B=B_\mathrm{max}$ and  black $B=B_\mathrm{min}$.
\label{FIG_HYBRID}}
\end{figure}

\section{Combining omnigenous and pwO fields}\label{SEC_HYBRIDS}

In this section, we briefly illustrate how the framework of this paper allows us to parametrize magnetic fields which behave as omnigenous fields (that can be described in terms of \cite{cary1997omni}) for deeply trapped particles and as pwO fields (as in section \ref{SEC_STRAIGHT}) for the rest. This is a straightforward way to allow $B$ to take a continuum of values in the range $[B_\mathrm{min}, B_0]$ outside the parallelogram parametrized in the previous sections. In practical terms, this could represent the result of slightly relaxing the constraints of \cite{cary1997omni} for barely trapped particles while keeping zero effective ripple. This can be of practical relevance as, for instance, quasi-isodynamicity cannot be achived exactly. When nearly QI fields are studied analytically, see e.g.~\cite{plunk2019direct,rodriguez2023qi2}, quasi-isodynamicity is typically broken for barely trapped particles.

In a QS field the $B$-contours must close in a direction given by constant $M\theta-N_pN\zeta$, i.e., they must have a helicity $(N,M)$, with integer $N$ and $M$. This is the only way that quasisymmetry,
\begin{eqnarray}
B(\zeta,\theta)=B^*(M\theta-N_pN\zeta),
\end{eqnarray}
is consistent with periodicity,
\begin{eqnarray}
B(\zeta+2\pi/N_p,\theta)=B(\zeta,\theta+2\pi)=B(\zeta,\theta),
\end{eqnarray}
and continuity at $\zeta=0$ and $\theta=0$. In this section, we will make deeply trapped particles behave as in a QS field by making the contours of $B < B_0$ close in the toroidal, poloidal or helical direction. Without loss of generality, we will make them straight lines, a simplification that will be lifted in section \ref{SEC_SMOOTHEST}.

In the case of poloidal symmetry, $M=0$, shown in figure \ref{FIG_HYBRID} (top left),
\begin{eqnarray}
B_{QPS+pwQS} =
\begin{cases}
B_{pwQS}(\zeta,\theta) & \text{if } \zeta_-\le\zeta\le\zeta_+, \\
B_{QPS}(\zeta) & \text{if } \zeta<\zeta_- \text{ or } \zeta>\zeta_+,
\end{cases}\label{EQ_QPSppwQS}
\end{eqnarray}
with $B_{pwQS}(\zeta,\theta)$ given by equation (\ref{EQ_BPWQS}) and
\begin{eqnarray}
B_{QPS} &< & B_{0},\\  
\lim_{\zeta\to\zeta_+} B_{QPS}&=&\lim_{\zeta\to\zeta_-} B_{QPS}=B_0,\\
\zeta_+ &=& \mathrm{max}(\zeta_{++},\zeta_{+-}),\\
\zeta_- &=& \mathrm{min}(\zeta_{-+},\zeta_{--}),\\
\zeta_-& +& \frac{2\pi}{N_p} >  \zeta_+.
\end{eqnarray}

In the case of toroidal symmetry, $N=0$, shown in figure \ref{FIG_HYBRID} (top right),
\begin{eqnarray}
B_{QAS+pwQS} =
\begin{cases}
B_{pwQS}(\zeta,\theta) & \text{if } \theta_-\le\theta\le\theta_+, \\
B_{QAS}(\theta) & \text{if } \theta<\theta_- \text{ or } \theta>\theta_+,
\end{cases}
\end{eqnarray}
with 
\begin{eqnarray}
B_{QAS} &<& B_{0},\\  
\lim_{\theta\to\theta_+} B_{QAS}&=&\lim_{\theta\to\theta_-} B_{QAS}=B_0,\\
\theta_+ &=& \mathrm{max}(\theta_{++},\theta_{-+}),\\
\theta_- &=& \mathrm{min}(\theta_{+-},\theta_{--}),\\
\theta_-& +& 2\pi >  \theta_+.
\end{eqnarray}

Finally, the helically symmetric case $M=1$, $N=-1$, shown in figure \ref{FIG_HYBRID} (bottomleft), reads
\begin{eqnarray}
B_{QHS(1,-1)+pwQS} =
\begin{cases}
  B_{pwQS}(\zeta,\theta) & \text{if }  \theta_{--}+N_p\zeta_{--}\le\theta+N_p\zeta\le \theta_{++}+N_p\zeta_{++}, \\
  & \text{or } \theta-\theta_c-t_2\zeta_c+2\pi+N_p\zeta<w_2 \\
  & \text{or } \theta-\theta_c-t_2\zeta_c-2\pi+N_p\zeta>-w_2 \\  
B_{QHS(1,-1)}(\theta+N_p\zeta) & \text{elsewhere},
\end{cases}
\end{eqnarray}
with
\begin{eqnarray}
B_{QHS(1,-1)} &<& B_{0},\\
\lim_{\theta+N_p\zeta\to\theta_{--} +N_p\zeta_{--}} B_{QHS(1,-1)} &=& \lim_{\theta+N_p\zeta\to\theta_{++} +N_p\zeta_{++}} B_{QHS(1,-1)} = \nonumber \\
\lim_{\theta+N_p\zeta\to \theta_c+t_2\zeta_c-2\pi+w_2} B_{QHS(1,-1)} &=& \lim_{\theta+N_p\zeta\to \theta_c+t_2\zeta_c+2\pi-w_2} B_{QHS(1,-1)} = B_0\\
\theta_{--}+N_p\zeta_{--}+2\pi&>&\theta_{++}+N_p\zeta_{++},
\end{eqnarray}
and the helically symmetric case $M=1$, $N=+1$, shown in figure \ref{FIG_HYBRID} (bottom right), reads
\begin{eqnarray}
B_{QHS(1,+1)+pwQS} =
\begin{cases}
B_{pwQS}(\zeta,\theta) & \text{if }  \theta_{+-}-N_p\zeta_{+-}\le\theta-N_p\zeta\le \theta_{-+}-N_p\zeta_{-+}, \\
  & \text{or } \theta-\theta_c-t_2\zeta_c+2\pi-N_p\zeta<w_2 \\
& \text{or } \theta-\theta_c-t_2\zeta_c-2\pi-N_p\zeta>-w_2 \\
B_{QHS(1,+1)}(\theta-N_p\zeta) & \text{elsewhere},
\end{cases}
\end{eqnarray}
with 
\begin{eqnarray}
B_{QHS)(1,+1)} &< & B_{0},\\  
\lim_{\theta-N_p\zeta\to\theta_{+-} +N_p\zeta_{+-}} B_{QHS(1,+1)} &=& \lim_{\theta-N_p\zeta\to\theta_{-+} +N_p\zeta_{-+}} B_{QHS(1,+1)} = \nonumber \\
\lim_{\theta-N_p\zeta\to \theta_c+t_2\zeta_c -2\pi+w_2} B_{QHS(1,+1)} &=& \lim_{\theta-N_p\zeta\to \theta_c+t_2\zeta_c+2\pi-w_2} B_{QHS(1,+1)} = B_0\\
\theta_{+-}-N_p\zeta_{+-}+2\pi&>&\theta_{-+}-N_p\zeta_{-+}.\label{EQ_CQ1}
\end{eqnarray}Examples with the four different helicities are shown in figure \ref{FIG_HYBRID}. For all particles that are trapped deeply enough, the magnetic field is practically QS, and no orbit-averaged radial drift exists. Less deeply trapped particles can be classified in three classes, as in section~\ref{SEC_STRAIGHT}. Within a class, the distance between bounce points does not depend on the field line, and neither does the distance in which $B$ is not constant. Since all trapped particles in a region see the same non-constant $B(\zeta-\zeta_{b_1},\theta)$ along the field line, the field is piecewise omnigenous.

Some cases are possible only for the right combinations of $t_1$, $t_2$, $w_1$ and $w_2$. For instance, positive $t_1$ and $t_2$ facilitate the case $M=1, N=-1$ and make the case $M=1, N=+1$ harder, as it becomes increasingly complicated to comply with equation (\ref{EQ_CQ1}). This can be compensated with a relatively small $w_1$ and/or $w_2$ (see figure \ref{FIG_HYBRID}, bottom left), which leaves more room for a region of helically closed $B$-contours. Similarly, larger helicities ($|M|>1$, $|N|>1)$ are possible but harder to achieve. We finally note that it may result natural to align two of the straight segments of $B=B_\mathrm{max}$ with the contours of $B=B_\mathrm{min}$ (as in figure \ref{FIG_HYBRID}, left), but this is not a requirement of piecewise omnigenity (see for instance figure \ref{FIG_HYBRID}, right).

Lastly, even if their appearance is smoother than those of section \ref{SEC_STRAIGHT}, these fields still have a discontinuity between the region with $B_\mathrm{min}\le B\le B_0$ and the region of $B=B_\mathrm{max}$. This is a requirement of piecewise omnigenity, as it will be illustrated again in section \ref{SEC_SCAN}.

\section{Other pwO fields with $B_\mathrm{min}\le B\le B_0<B_\mathrm{max}$ or $B=B_\mathrm{max}$}\label{SEC_SCAN}

The non smoothness of $B$ in pwO fields of section~\ref{SEC_STRAIGHT} has been alleviated to some extent by combining omnigenity and piecewise omnigenity, as in section \ref{SEC_HYBRIDS}. In this section we present an alternative way of achieving this, by constinuously modifing (a subset of) the fields of section~\ref{SEC_STRAIGHT}. The connection between pwO fields and omnigenous fields will appear more clearly, as well as the role of the discontinuity in $B$.

We impose the following conditions on a smooth pwO field, $B_{spwO}(\zeta,\theta)$:
\begin{enumerate}
\item In the area contained by a parallelogram, $B=B_\mathrm{max}$.
\item The rotational transform is such that only two field lines connect the four corners of this parallelogram.
\item In the region of $B<B_\mathrm{max}$ enclosed by these two field lines and the two sides of the parallelogram connected by them, the contours of constant $B$ are parallel to these sides.
\item On the rest of the flux-surface, the contours of constant $B$ are parallel to the other two sides of the parallelogram.
\item $B$ is discontinuous at the sides of the parallelogram and continuous elsewhere on the flux-surface.
\end{enumerate}

\begin{figure}
\includegraphics[angle=0,width=\columnwidth]{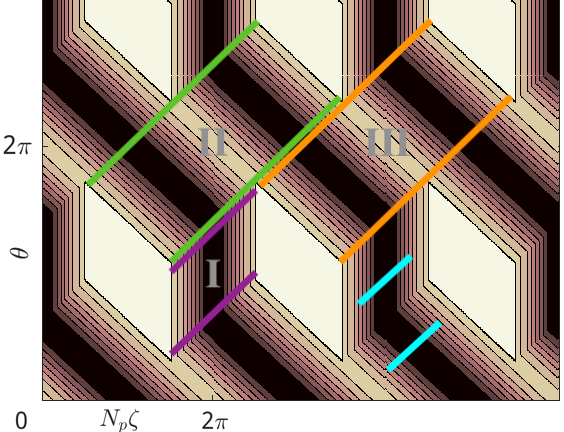}  
\caption{$B_{spwQS}$ of equations (\ref{EQ_SBPWQS}), (\ref{EQ_IOTA0}), (\ref{EQ_CONTIII}) and (\ref{EQ_DISC}). All parameters are the same of figure \ref{FIG_PWQS} except for $w_2$, which is set by equation (\ref{FIG_CONDW2}).
\label{FIG_sPWQS}}
\end{figure}

An example of such field is shown in figure \ref{FIG_sPWQS}. In the notation of section \ref{SEC_STRAIGHT},

\begin{eqnarray}
B_{spwQS}(\zeta,\theta)=
\begin{cases}
  B_\mathrm{max}  & \text{if } \zeta-\zeta_c+t_1(\theta-\theta_c)\ge -w_1  \\ 
                             & \text{and } \zeta-\zeta_c+t_1(\theta-\theta_c)\le +w_1 \\ 
                             & \text{and } \theta-\theta_c+t_2(\zeta-\zeta_c)\ge -w_2 \\
                             & \text{and } \theta-\theta_c+t_2(\zeta-\zeta_c)\le +w_2, \\

B_I(\zeta,\theta)=B_I^*(\zeta+t_1\theta)  & \text{if } \zeta-\zeta_c+t_1(\theta-\theta_c)>+w_1 \\ 
B_I(\zeta,\theta)<B_\mathrm{max}    & \text{and } \zeta-\zeta_c+t_1(\theta-\theta_c)<-w_1+2\pi/N_p \\ 
                             & \text{and } \theta\le\theta_{++}+\iota_0(\zeta-\zeta_{++})\\ 
                      & \text{and } \theta\ge\theta_{+-}+\iota_0(\zeta-\zeta_{+-}),\\ 
B_{II}(\zeta,\theta)=B_{II}^*(\theta+t_2\zeta)<B_\mathrm{max}  & \text{elsewhere,}
\end{cases}\label{EQ_SBPWQS}
\end{eqnarray}
complemented with equation (\ref{EQ_IOTA0}). Continuity between the regions of $B_{spqQS}=B_{I}$ and $B_{spqQS}=B_{II}$ requires
\begin{eqnarray}
B_I(\zeta,\theta_{++}+\iota(\zeta-\zeta_{++}))=B_{II}(\zeta,\theta_{++}+\iota(\zeta-\zeta_{++})),\label{EQ_CONTIII}
\end{eqnarray}
and discontinuity at the segments where $B_{spqQS}=B_\mathrm{max}$ means
\begin{eqnarray}
\lim_{\zeta-\zeta_c+t_1(\theta-\theta_c)\to w_1}B_I=\lim_{\theta-\theta_c+t_2(\zeta-\zeta_c)\to w_2}B_{II}<B_\mathrm{max}.\label{EQ_DISC}
\end{eqnarray}

It can be checked that the previous conditions constrain the shape of possible parallelograms: continuity between the regions outside the parallelogram, $B_{I}(\zeta,\theta)=B_{I}^*(\zeta+t_1\theta)$ and $B_{II}(\zeta,\theta)=B_{II}^*(\theta+t_2\zeta)$, can be fulfilled non-trivially if lines
\begin{eqnarray}
\theta-\theta_c+t_2(\zeta-\zeta_c-2k\pi/N_p)= +w_2
\end{eqnarray}
and
\begin{eqnarray}
 (\theta-\theta_c-2\pi)+t_2(\zeta-\zeta_c)= -w_2
\end{eqnarray}
coincide, with $k=1,2...$, which implies
\begin{eqnarray}
 w_2=\pi\left(1-\frac{kt_2}{N_p}\right).\label{FIG_CONDW2}
\end{eqnarray}
Other relations between the parameters, for other choices of $\iota$ in section~\ref{SEC_ALTIOTA}, are possible. For a generic choice of parameters $t_1$, $t_2$, $w_1$ and $w_2$, the contours of $B< B_\mathrm{max}$ must fill the area outside the parallelogram, i.e., $B_I=B_{II}=B_0<B_\mathrm{max}$, and we recover the field of equation (\ref{EQ_BPWQS}).

Figure \ref{FIG_sPWQS} follows equation (\ref{FIG_CONDW2}). The contours of $B<B_0$ close over themselves without local maxima, but do not follow lines of constant $M\theta-N_pN\zeta$: they are consisting of segments that follow either constant $\zeta+t_1\theta$ or constant $\theta+t_2\zeta$. It is straightforward to see that figure \ref{FIG_sPWQS} generalizes the field of figure \ref{FIG_PWQS} while still fulfilling equations (\ref{EQ_PWO}) and (\ref{EQ_TRANSITIONS}): for a range of values of $\mathcal{E}/\mu$ there exist several classes of trapped particles, particles transition between regions and, within a given region, the distance between bounce points and the value of $B$ along the field line does not depend on the field line. For values of $\mathcal{E}/\mu$ close enough to $B_\mathrm{min}$, only one class of trapped particles exist (light blue). They behave as in a QS field of helicity $(1,-t_2/N_p)$ or $(t_1,-1/N_p)$, depending on the field line.

\begin{figure}
\includegraphics[angle=0,width=\columnwidth]{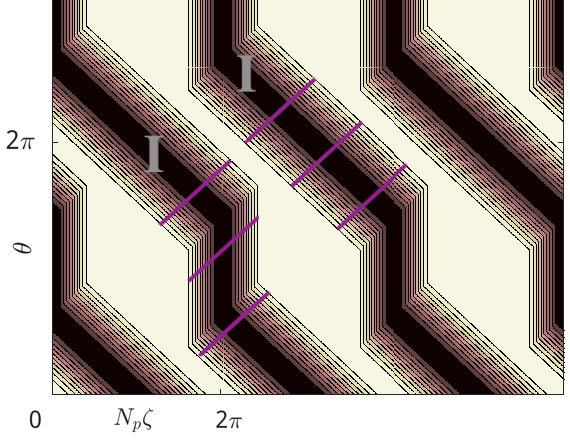}  
\caption{$B_{spwQS}$ of equations (\ref{EQ_SBPWQS}), (\ref{EQ_IOTA0}), (\ref{EQ_CONTIII}) and (\ref{EQ_CONT}). All parameters are the same of figure \ref{FIG_sPWQS}.\label{FIG_OH}}
\end{figure}

\begin{figure}
\includegraphics[angle=0,width=0.49\columnwidth]{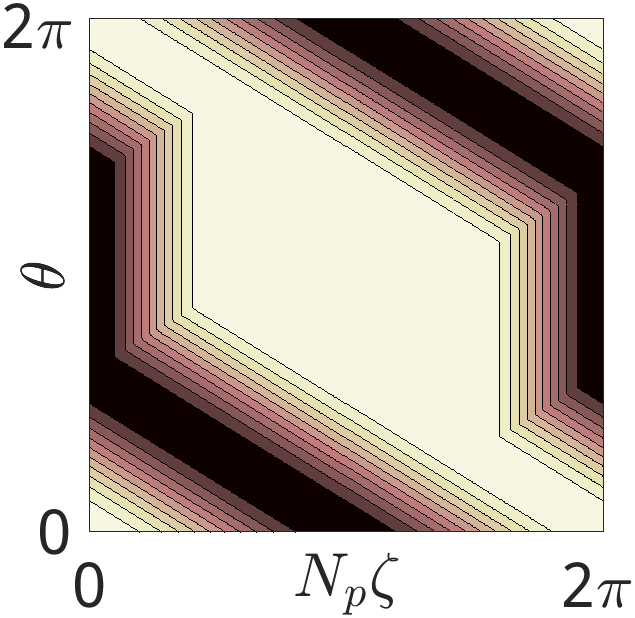}
\includegraphics[angle=0,width=0.49\columnwidth]{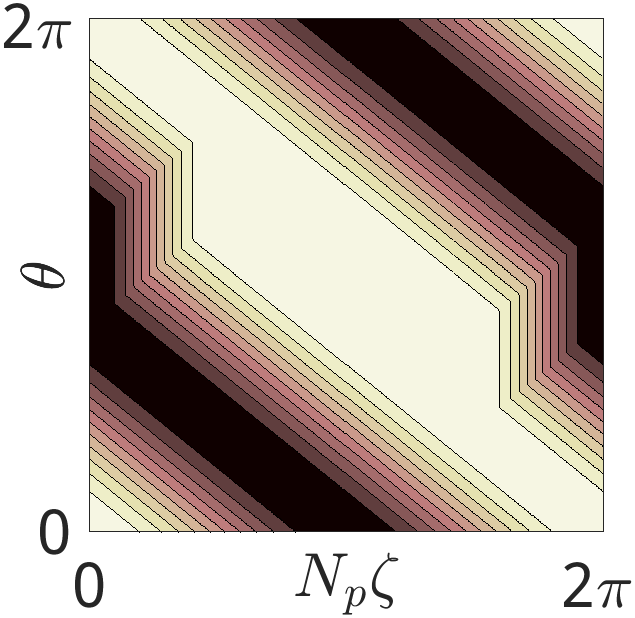}
\includegraphics[angle=0,width=0.49\columnwidth]{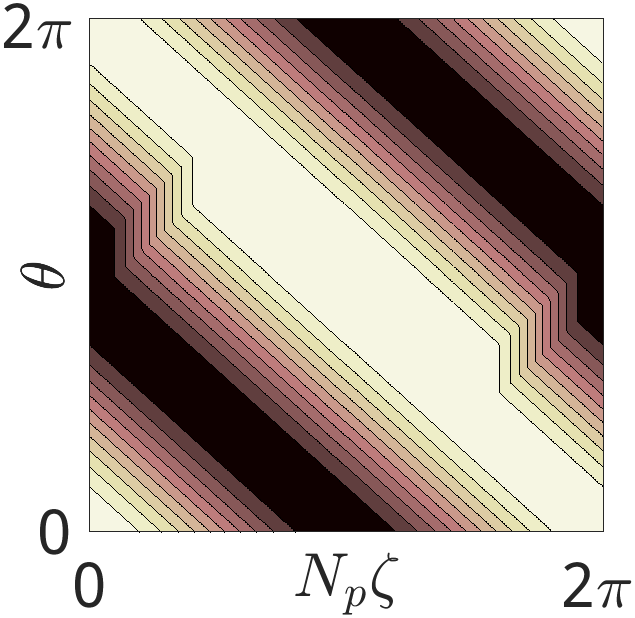}
\includegraphics[angle=0,width=0.49\columnwidth]{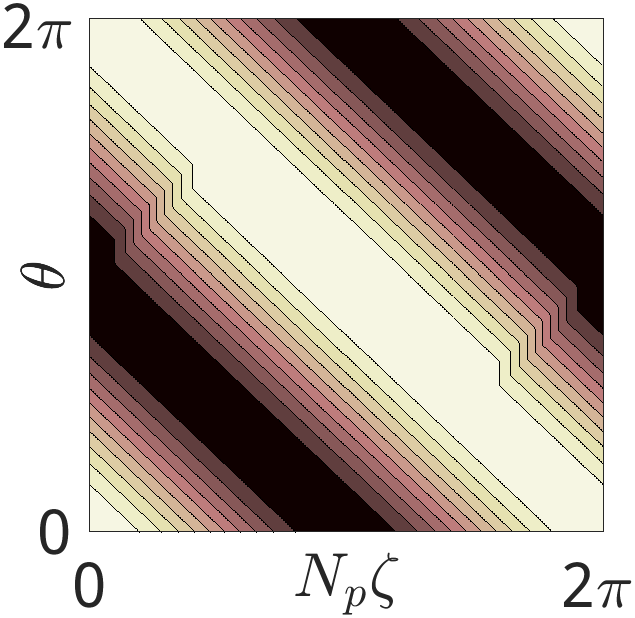}  
\caption{$B_{spwQS}$ of figure \ref{FIG_OH} with same $w_2$ (top left), with $w_2$ divided by 2 (top right), by 4 (bottom left) and by 8 (bottom right).
\label{FIG_SCAN}}
\end{figure}

We end this section by illustrating that the discontinuity in $B_{spwQS}(\zeta,\theta)$ is unavoidable. Let us assume that
\begin{eqnarray}
\lim_{\zeta-\zeta_c+t_1(\theta-\theta_c)\to w_1}B_I=\lim_{\theta-\theta_c+t_2(\zeta-\zeta_c)\to w_2}B_{II}=B_\mathrm{max}.\label{EQ_CONT}
\end{eqnarray}
Then, the field of figure \ref{FIG_sPWQS} changes to the one in figure \ref{FIG_OH}. The $B_\mathrm{max}$ contours now close helically (even if they are still consisting of segments constant $\zeta+t_1\theta$ or constant $\theta+t_2\zeta$ rather than of constant $M\theta-N_pN\zeta$). There is only one class of trapped particles, and the field can therefore be considered omnigenous as in \cite{cary1997omni}, rather than as piecewise omnigenous. The difference with the other omnigenous fields presented in the literature so far is that $B=B_\mathrm{max}$ does not lie on a line (as explicitely assumed in~\cite{cary1997omni}) or on a discrete number of lines (as in \cite{parra2015omni}), but on an area of the flux-surface. While these fields break analiticity at a lower order (the first derivative of $B$ is discontinuous), this does not rule out that they could be easier to approach in stellarator optimization.






The effect of varying this area (through $w_2$, which in turn requires modifying $t_2$ and $\iota$) can be seen in figure \ref{FIG_SCAN}. As $w_2$ decreases and $t_2$ changes accordingly towards the integer value $N_p$, the field becomes closer to quasisymmetric with $M=1$, $N=-1$. Conversely, we learn that the omnigenity condition that the $B_\mathrm{max}$-contour must be a straight line of constant $M\theta-N_pN\zeta$ \cite{cary1997omni,parra2015omni} can be relaxed by making the region $B=B_\mathrm{max}$ adopt the shape of a parallelogram. The size of this region determines how far can the field be from depending on the Boozer angles through $M\theta-N_pN\zeta$.

We end this section with another example field constructed according to equation (\ref{FIG_CONDW2}). This one resembles W7-X standard, as it can be seen by comparing with figure (\ref{FIG_TRANSITION}) left, but its transitioning particles will fulfill piecewise omnigenity.

\begin{figure}
\includegraphics[angle=0,width=0.49\columnwidth]{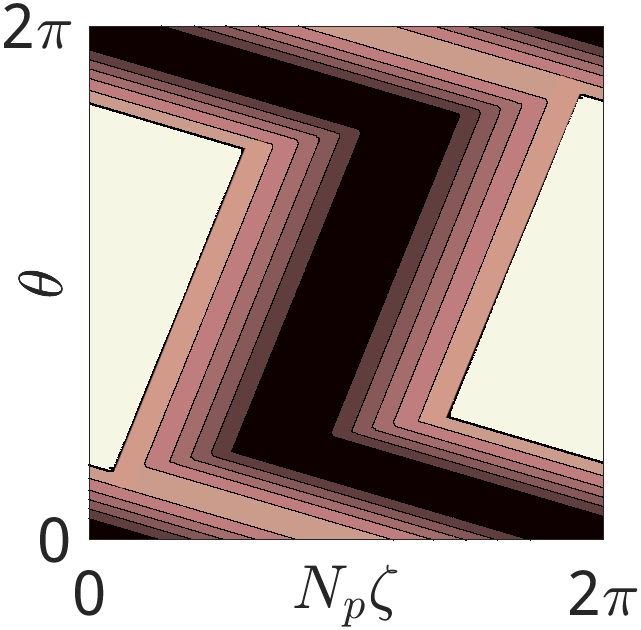}
\caption{$B_{spwQS}$ that resembles the field of W7-X standard.
\label{FIG_W7-XlikePWO}}
\end{figure}

\begin{figure}
\includegraphics[angle=0,width=0.49\columnwidth]{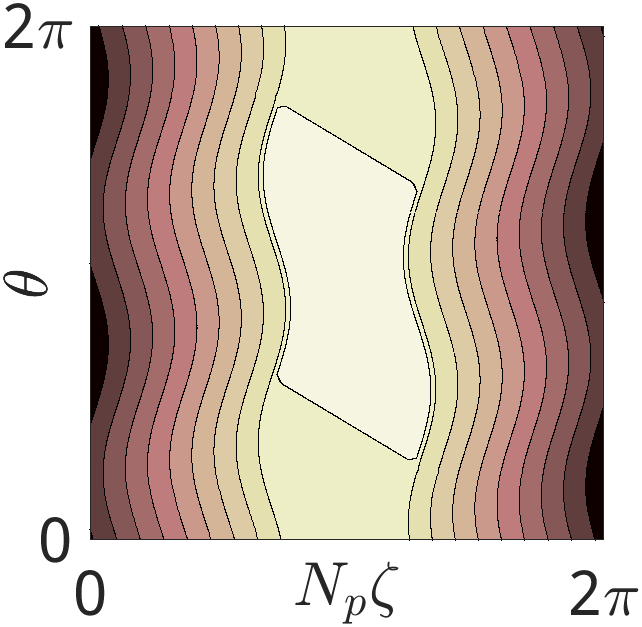}  
\includegraphics[angle=0,width=0.49\columnwidth]{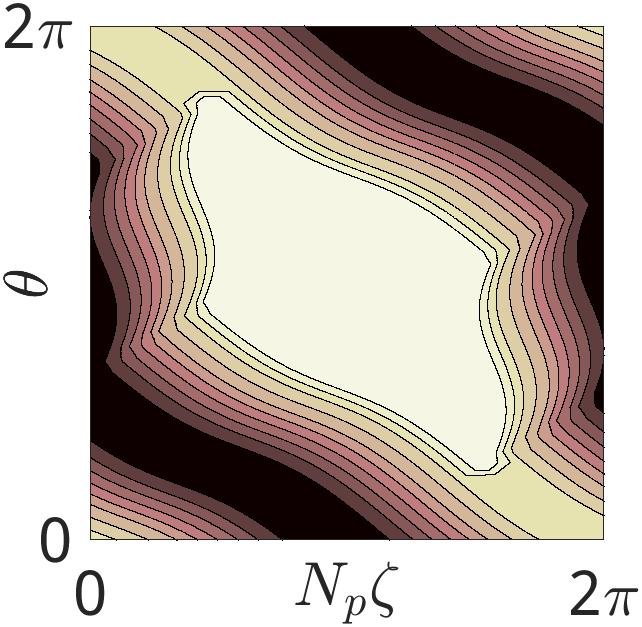}  
\caption{pwO fields with $B$-contours consisting of non-straight segments and $B_\mathrm{min}\le B\le B_0<B_\mathrm{max}$ or $B=B_\mathrm{max}$\label{FIG_SMOOTHEST}}
\end{figure}

\section{Smooth piecewise omnigenous fields with $B$-contours consisting of non straight segments}\label{SEC_SMOOTHEST}

In this section,we combine the results of the previous sections. In particular, we use the lessons learnt in section \ref{SEC_NONSTRAIGHT} to generalize the fields of sections \ref{SEC_HYBRIDS} and \ref{SEC_SCAN}.

We start with section \ref{SEC_HYBRIDS}. We aim at fields that are piecewise omnigenous with non-straigth contours for barely trapped particles and non-quasisymmetric omnigenous for deeply trapped particles. Without loss of generality, we will focus on approximately QI fields: deeply trapped particles will see $B$-contours that close poloidally. This means that we will be generalizing equation (\ref{EQ_QPSppwQS}). We propose
\begin{eqnarray}
B_{QI+pwO} =
\begin{cases}
B_{pwO}(\zeta,\theta) & \text{if } \zeta_-(\theta)\le\zeta\le\zeta_+(\theta), \\
B_{QI}(\zeta,\theta) & \text{if } \zeta<\zeta_-(\theta) \text{ or } \zeta>\zeta_+(\theta),
\end{cases}\label{EQ_QIppwQS}
\end{eqnarray}
Here, $B_{QI}(\zeta,\theta)$ may be any field that is quasi-isodynamic for $\iota=\iota_0$ such that its maximum value lies at $\zeta=\pi/N_p$ and for which $\zeta_-(\theta)$ and $\zeta_+(\theta)$ are contours of constant field strength $B_{QI}=B_0$. The pwO piece, $B_{pwO}(\zeta,\theta)$, is such such that 
\begin{eqnarray}
B_{pwO}(\zeta=\zeta_-(\theta))=B_{pwO}(\zeta=\zeta_+(\theta))=B_0
\end{eqnarray}
A simple choice would be $\zeta_-(\theta)$ and $\zeta_+(\theta)$ to be straight lines, as in equation (\ref{EQ_QPSppwQS}). A smoother choice is to make $\zeta_+(\theta)$ and $\zeta_-(\theta)$ contain the contour 
\begin{eqnarray}
\zeta-\zeta_c+t_1(\theta-\theta_c)=+w_1+\delta w_1
\end{eqnarray}
between $\zeta_{++}$ and $\zeta_{+-}$ and the contour 
\begin{eqnarray}
\zeta-\zeta_c+t_1(\theta-\theta_c)=-w_1+\delta w_1
\end{eqnarray}
between $\zeta_{-+}$ and $\zeta_{--}$. We do not devote time to parametrization of $B_{QI}$, since this has already been done e.g. in \cite{dudt2023omni}. Figure~\ref{FIG_SMOOTHEST} (left) shows an example that generalizes figure~\ref{FIG_HYBRID} (top left).

We next undertake generalization of the fields of section \ref{SEC_SCAN}. We propose, as generalization of equation (\ref{EQ_SBPWQS}),  
\begin{eqnarray}
B_{spwO}(\zeta,\theta)=
\begin{cases}
  B_\mathrm{max}  & \text{if } \zeta-\zeta_c+t_1(\theta-\theta_c)\ge -w_1+\delta w_1  \\ 
                             & \text{and } \zeta-\zeta_c+t_1(\theta-\theta_c)\le +w_1+\delta w_1 \\ 
                             & \text{and } \theta-\theta_c+t_2(\zeta-\zeta_c)\ge -w_2 +\delta w_2\\
                             & \text{and } \theta-\theta_c+t_2(\zeta-\zeta_c)\le +w_2+\delta w_2, \\

B_I(\zeta,\theta)=B_{\eta_I}(\eta_I)  & \text{if } \zeta-\zeta_c+t_1(\theta-\theta_c)>+w_1+\delta w_1 \\ 
B_I(\zeta,\theta)<B_\mathrm{max}    & \text{and } \zeta-\zeta_c+t_1(\theta-\theta_c)<-w_1+\delta w_1+2\pi/N_p \\ 
                             & \text{and } \theta\le\theta_{++}+\iota_0(\zeta-\zeta_{++})\\ 
                      & \text{and } \theta\ge\theta_{+-}+\iota_0(\zeta-\zeta_{+-}),\\ 
B_{II}(\zeta,\theta)=B_{\eta_{II}}(\eta_{II})<B_\mathrm{max}  & \text{elsewhere,}
\end{cases}\label{EQ_SBPWO}
\end{eqnarray}
complemented with equation (\ref{EQ_IOTA0}). Function $\delta w_1$ is defined in equations (\ref{EQ_SIDEPI}) and (\ref{EQ_SIDEPII}), and function $\delta w_2$ in equations (\ref{EQ_SIDEMI}) and (\ref{EQ_SIDEMII}). 
Discontinuity at the segments where $B_{spwO}=B_\mathrm{max}$ means
\begin{eqnarray}
\lim_{\zeta-\zeta_c+t_1(\theta-\theta_c)\to w_1+\delta w_1}B_{\eta_I}=\lim_{\theta-\theta_c+t_2(\zeta-\zeta_c)\to w_2+\delta w_2}B_{\eta_{II}}<B_\mathrm{max}.
\end{eqnarray}
Omnigenity within regions I and II (and continuity between them) is achieved through constraints in functions $B_{\eta_\mathrm{w}}$ and $\eta_\mathrm{w} $, with w=I, II. This is equivalent to the Cary and Shasharina construction (restricted to an angular region of the field period). We first define the auxiliar coordinates
\begin{eqnarray}
\theta_\mathrm{I} &=& 2\pi\frac{\theta-\iota\zeta-(\theta_{+-}-\iota\zeta_{+-})}{\theta_{++}-\iota\zeta_{++}-(\theta_{+-}-\iota\zeta_{+-})},\\
\zeta_\mathrm{I} &=& 2\pi\frac{\zeta-\zeta_{b_1}(\zeta,\theta)}{\zeta_{b_2}(\zeta,\theta)-\zeta_{b_1}(\zeta,\theta)},
\end{eqnarray}
in region I, and
\begin{eqnarray}
\theta_\mathrm{II} &=& 2\pi\frac{\theta-\iota\zeta-(\theta_{++}-\iota\zeta_{++})}{\theta_{-+}-\iota\zeta_{-+}-(\theta_{++}-\iota\zeta_{++})},\\
\zeta_\mathrm{II} &=& 2\pi\frac{\zeta-\zeta_{b_1}(\zeta,\theta)}{\zeta_{b_2}(\zeta,\theta)-\zeta_{b_1}(\zeta,\theta)},
\end{eqnarray}
in region II. There, $\zeta_{b_1}(\zeta,\theta)$ and $\zeta_{b_2}(\zeta,\theta)$ are the bounce points for a trajectory with ${\mathcal{E}}/\mu=B_\mathrm{max}$ that goes through point $(\zeta,\theta)$. Omnigenity within regions I and II (and continuity between them) is achieved with an adequate choice of functions $\eta_\mathrm{I}(\zeta_\mathrm{I},\theta_\mathrm{I})$ and $\eta_\mathrm{II}(\zeta_\mathrm{II},\theta_\mathrm{II})$. The shape of such functions was discussed in  \cite{cary1997omni,parra2015omni,landreman2012omni} and parametrized in \cite{dudt2023omni}. The trivial choice, $\eta_\mathrm{w}=\zeta_\mathrm{w}$ and $B_{\eta_\mathrm{I}}=B_{\eta_\mathrm{II}}$, is shown in figure~\ref{FIG_SMOOTHEST} (right).

\begin{figure}
\includegraphics[angle=0,width=0.49\columnwidth]{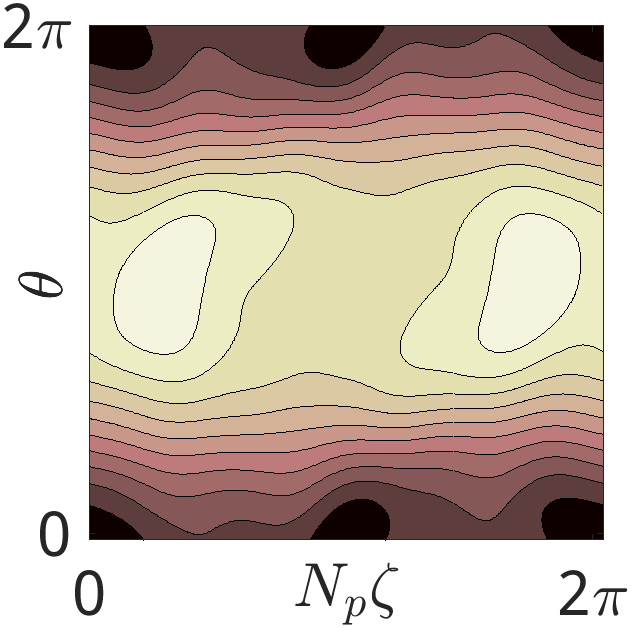}
\includegraphics[angle=0,width=0.49\columnwidth]{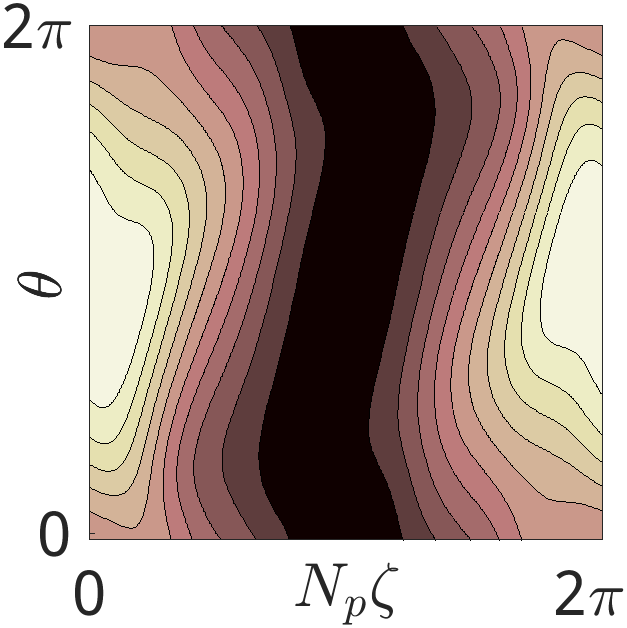}
\caption{$B$ on flux-surface $s=0.25$ of NCSX (left), an equilibrium in the configuration space of CIEMAT-QI4 (right).
\label{FIG_EXAMPLES}}
\end{figure}


\section{Discussion}\label{SEC_SUMMARY}

Piecewise omnigenous fields, recently introduced in~\cite{velasco2024pwO}, are stellarator magnetic fields that are optimized with respect to radial neoclassical transport. They are qualitatively different from omnigenous fields, a family that includes quasi-isodynamic and quasisymmetric fields, and this paper has been a first attemp to characterize and parametrize them in a systematic manner.

In order for the concept of piecewise omnigenity to be relevant, it is required that magnetohydrodynamic (MHD) equilibria exist close enough to exact piecewise omnigenity. As a first rigorous assessment, closeness to omnigenity was qualitatively inferred in~\cite{velasco2024pwO} for reactor-relevant configurations by means of calculations of the second adiabatic invariant. In this work, we also tentatively employ visual inspection of $B(\theta,\zeta)$, very often a relevant part of the stellarator optimization process. As discussed in section~\ref{SEC_SCAN}, the magnetic field of W7-X standard, see figure~\ref{FIG_TRANSITION} (left), looks similar to the pwO field of figure \ref{FIG_W7-XlikePWO}. This could be part of the explanation why neoclassical transport is actually lower in this configuration than in the high-mirror configuration of W7-X, which is closer to being quasi-isodynamic~\cite{beidler2011ICNTS}.  In figure~\ref{FIG_EXAMPLES} (top right), we show NCSX, a compact quasi-axisimmetric configuration, which resembles the field of figure~\ref{FIG_HYBRID} (top right). This could partly explain why, even being clearly further from exact quasisymmetry than the stellarator HSX (see e.g.~\cite{beidler2011ICNTS}, figures 5 and 6), its effective ripple is smaller. Figure~\ref{FIG_EXAMPLES} (bottom left) shows a configuration obtained during the optimization campaign of CIEMAT-QI4~\cite{sanchez2023qi} that looks similar to figure~\ref{FIG_HYBRID} (top left) and has a smaller effective ripple than CIEMAT-QI4 and reduced elongation. Finally, a magnetic configuration that is QI for deeply trapped particles and pwO for barely trapped particles, similarly to figure~\ref{FIG_HYBRID} (top left), has been reported in~\cite{liu2024omni}. Interestingly, it displays reduced flux-surface shaping with respect to what it is usually the case in QI configurations, see~\cite{rodriguez2024maxj} and references therein.

The discussion of the previous paragraph merely illustrates the enormous size of the region of the stellarator configuration space that is close to being pwO. In this vast region, one should expect to find configurations with varying degree of, for instance, elongation, compacity, or turbulent transport level. There could also exist families of nearly pwO configurations that are more appropriate for one or another divertor concept (the identification pwO fields with a reduced bootstrap current, and thus compatible in principle with an island divertor, is underway). It is noteworthy that some of the nearly pwO configurations mentioned in this section have been obtained by simply not imposing omnigenity for a subset of particle orbits during optimization~\cite{liu2024omni} or by adding other reactor-relevant design criteria to omnigenity during the optimization process. This leads us to expect that the exploration of the space of nearly pwO configurations will naturally produce novel and relevant reactor candidates, and the work in this paper constitutes the first systematic step in this direction.

\begin{acknowledgments}
This work has been carried out within the framework of the EUROfusion Consortium, funded by the European Union via the Euratom Research and Training Programme (Grant Agreement No 101052200 - EUROfusion). Views and opinions expressed are however those of the author(s) only and do not necessarily reflect those of the European Union or the European Commission. Neither the European Union nor the European Commission can be held responsible for them. This research was supported in part by Grant No. PID2021-123175NB-I00, funded by Ministerio de Ciencia, Innovaci\'on y Universidades / Agencia Estatal de Investigaci\'on / 10.13039/501100011033 and by ERDF/EU. The authors acknowledge useful discussions with E. Rodr\'iguez, J. Escoto and EUROfusion's TSVV12 team.

 \end{acknowledgments}

\bibliography{/Users/velasco/Work/PAPERS/bibliography.bib}

\end{document}